\PassOptionsToPackage{unicode}{hyperref}
\PassOptionsToPackage{hyphens}{url}
\documentclass[
  11pt,
]{article}
\usepackage{amsmath,amssymb}
\usepackage{lmodern}
\usepackage{iftex}
\ifPDFTeX
  \usepackage[T1]{fontenc}
  \usepackage[utf8]{inputenc}
  \usepackage{textcomp} % provide euro and other symbols
\else % if luatex or xetex
  \usepackage{unicode-math}
  \defaultfontfeatures{Scale=MatchLowercase}
  \defaultfontfeatures[\rmfamily]{Ligatures=TeX,Scale=1}
\fi
\IfFileExists{upquote.sty}{\usepackage{upquote}}{}
\IfFileExists{microtype.sty}{% use microtype if available
  \usepackage[]{microtype}
  \UseMicrotypeSet[protrusion]{basicmath} % disable protrusion for tt fonts
}{}
\makeatletter
\@ifundefined{KOMAClassName}{% if non-KOMA class
  \IfFileExists{parskip.sty}{%
    \usepackage{parskip}
  }{% else
    \setlength{\parindent}{0pt}
    \setlength{\parskip}{6pt plus 2pt minus 1pt}}
}{% if KOMA class
  \KOMAoptions{parskip=half}}
\makeatother
\usepackage{xcolor}
\IfFileExists{xurl.sty}{\usepackage{xurl}}{} % add URL line breaks if available
\IfFileExists{bookmark.sty}{\usepackage{bookmark}}{\usepackage{hyperref}}
\hypersetup{
  pdftitle={Estimating causes of maternal death in data-sparse contexts},
  pdfauthor={Monica Alexander; Michael Y.C. Chong; Marija Pejcinovska},
  hidelinks,
  pdfcreator={LaTeX via pandoc}}
\usepackage[margin=1in]{geometry}
\usepackage{graphicx}
\makeatletter
\def\maxwidth{\ifdim\Gin@nat@width>\linewidth\linewidth\else\Gin@nat@width\fi}
\def\maxheight{\ifdim\Gin@nat@height>\textheight\textheight\else\Gin@nat@height\fi}
\makeatother
\setkeys{Gin}{width=\maxwidth,height=\maxheight,keepaspectratio}
\makeatletter
\def\fps@figure{htbp}
\makeatother
\providecommand{\tightlist}{%
  \setlength{\itemsep}{0pt}\setlength{\parskip}{0pt}}
\usepackage{setspace}
\usepackage{tikz}
\usetikzlibrary{arrows}
\definecolor{OliveGreen}{RGB}{34,139,34}
\usepackage{booktabs}
\usepackage{longtable}
\usepackage{booktabs}
\usepackage{longtable}
\usepackage{array}
\usepackage{multirow}
\usepackage{wrapfig}
\usepackage{float}
\usepackage{colortbl}
\usepackage{pdflscape}
\usepackage{tabu}
\usepackage{threeparttable}
\usepackage{threeparttablex}
\usepackage[normalem]{ulem}
\usepackage{makecell}
\usepackage{xcolor}
\ifLuaTeX
  \usepackage{selnolig}  % disable illegal ligatures
\fi
\newlength{\cslhangindent}
\newlength{\csllabelwidth}
\newenvironment{CSLReferences}[2] % #1 hanging-ident, #2 entry spacing
 {% don't indent paragraphs
  \setlength{\parindent}{0pt}
  \ifodd #1 \everypar{\setlength{\hangindent}{\cslhangindent}}\ignorespaces\fi
  \ifnum #2 > 0
  \setlength{\parskip}{#2\baselineskip}
  \fi
 }%
 {}
\usepackage{calc}

\title{Estimating causes of maternal death in data-sparse contexts}
\author{Monica Alexander\footnote{\href{mailto:monica.alexander@utoronto.ca}{\nolinkurl{monica.alexander@utoronto.ca}}.} \and Michael
Y.C. Chong \and Marija Pejcinovska}
\date{Department of Statistical Sciences, University of Toronto}

\begin{document}
\maketitle
\begin{abstract}
Understanding the underlying causes of maternal death across all regions
of the world is essential to inform policies and resource allocation to
reduce the mortality burden. However, in many countries there exists
very little data on the causes of maternal death, and data that do exist
do not capture the entire population at risk. In this paper, we present
a Bayesian hierarchical multinomial model to estimate maternal cause of
death distributions globally, regionally, and for all countries
worldwide. The framework combines data from various sources to inform
estimates, including data from civil registration and vital systems,
smaller-scale surveys and studies, and high-quality data from
confidential enquiries and surveillance systems. The framework accounts
for varying data quality and coverage, and allows for situations where
one or more causes of death are missing. We illustrate the results of
the model on three case-study countries that have different data
availability situations.
\end{abstract}

\newcommand{\HIV}{\text{HIV}}
\newcommand{\IND}{\text{IND}}
\newcommand{\DIR}{\text{DIR}}
\newcommand{\HEM}{\text{HEM}}
\newcommand{\SEP}{\text{SEP}}

\newpage

\hypertarget{introduction-and-motivation}{%
\section{Introduction and
Motivation}\label{introduction-and-motivation}}

Maternal mortality is an important indicator of the health and
development of a country. It is explicitly linked to child health and
development outcomes, and is strongly associated with women's education,
rights and access to adequate healthcare (McAlister and Baskett 2006;
Alvarez et al. 2009; Muldoon et al. 2011; Mbizvo and Say 2012; Alkema et
al. 2016; Briozzo et al. 2016). Reducing maternal deaths is a public
health priority, with Sustainable Development Goal (SDG) 3.1 aiming to
reduce the global maternal mortality ratio (MMR) to less than 70 deaths
per 100,000 live births by the year 2030 (UNDESA 2020). Over the past
several decades, substantial progress towards this goal has been made,
with the global estimate of maternal deaths in 2017 being 35\% lower
than deaths in 2000 (295,000 compared to 451,000) (WHO et al. 2019).
However, substantial disparities exist across countries; in particular,
Sub-Saharan African countries account for around 66\% of all maternal
deaths worldwide.

An important part of understanding differences in levels and trends in
overall mortality is determining the main causes of maternal death.
Understanding the main causes of death would help to guide how best to
improve conditions in regions of the world that are making relatively
slow progress towards the 2030 goals, offering direct insight into how
resources could be most effectively allocated to reduce overall
mortality. As such, the aim of this project was to estimate, for all
regions in the world, the proportion of maternal deaths due to major
causes, including hemorrhage, sepsis, hypertension, embolism, abortion,
indirect causes and other direct causes.

If data existed on every maternal death and the underlying and
associated causes, then it would be a simple tabulation exercise to
calculate the maternal cause of death distributions for regions
worldwide. However, as is the case with most other population statistics
and indicators, the degree of data availability on maternal deaths, and
the quality of such data, varies substantially across countries. One of
the main sources of data on deaths are civil registration and vital
statistics (CRVS) systems, which aim to record all vital events (such as
births and deaths) for a population. While well-functioning CRVS systems
exist in most high-income countries, many lower- and middle-income
countries (LMIC) have no such systems in place. Indeed, only 26\% of the
world's population lives in countries with complete registration of
deaths (CDC 2015). Countries that lack CRVS systems are concentrated in
regions where the overall mortality burden is highest, such as
Sub-Saharan Africa and Southern Asia. In such countries, other data
sources on maternal mortality may be available, such as data from
surveys, or local-level administrative records from hospitals or
community health centers. Such data are generally of lower quality and
unlikely to be representative of the broader population of interest.

Data issues are further exacerbated when considering the estimation of
causes of death. Even if data exist on the total number of maternal
deaths in a population, we may not have information on some or all of
the causes of maternal death. In a particular population or country, the
causes of maternal death reported may vary by year or data source. In
addition, the quality of reporting of causes of death varies
substantially across populations, even within CRVS systems, and is
dependent on factors such as country-level practices and policies, and
physician and health official training (Messite and Stellman 1996;
Salanave et al. 1999; Eriksson et al. 2013; MacDorman et al. 2016).
While cause of death information has a standardized classification
system --- the International Classification of Diseases (ICD), currently
in its 10th revision (ICD-10) --- it is not always the case that deaths
are reported according to ICD-10 classifications, particularly when data
are sourced from surveys or single institution studies.

Data quality and sparsity issues are particularly pertinent in the case
of maternal deaths compared to deaths at other ages (for example, child
mortality). Although the MMR is much higher than is acceptable in many
countries, the absolute number of maternal deaths relative to the number
of live births (the measure of exposure to risk) is relatively small. In
addition, an important cause of maternal death --- deaths due to
abortion --- are substantially under-reported in many countries, due to
definition, cultural and legality issues (Gerdts, Vohra, and Ahern 2013;
Gerdts et al. 2015; Abouchadi, Zhang, and De Brouwere 2018). All these
data characteristics make the goal of estimating a maternal cause of
death distribution particularly challenging.

In this paper we present a Bayesian hierarchical modeling framework to
estimate maternal cause of death distributions in contexts of varying
data availability and quality. The model estimates a set 14
sub-categories within 7 main categories of maternal death, for each
country of interest. The model accounts for varying data quality and
coverage, thereby allowing for many different types of data sources to
be included to inform estimates. The modeling framework can be easily
adapted to estimate maternal cause of death distributions in a variety
of different populations and time periods, in varying data availability
contexts.

The remainder of the paper is structured as follows. We first briefly
discuss the use of Bayesian modeling in global health estimation in
general, and specific efforts to model causes of death. We then outline
the goals of estimation and introduce the 14 different categories that
make up the cause of death distributions. Section \ref{section_data}
outlines the data sources and availability and how these were processed
into the main cause of death categories. Section \ref{section_model}
describes in detail the modeling framework and computational aspects. We
then illustrate some key features of the model through illustrative
results, and present results from validation exercises in Section
\ref{section_results}. Finally, we discuss possible extensions to the
modeling framework and future work in Section \ref{section_discussion}.

\hypertarget{background}{%
\section{Background}\label{background}}

Statistical modeling frameworks used in demographic and global health
estimation have increasingly shifted to using Bayesian methods over the
past decade. The introduction of Bayesian methods by the United Nations
Population Division to estimate world population trends from 2010
(Raftery, Alkema, and Gerland 2014) paved the way for other UN
organizations to follow suit. Many of the global health indicators
related to mortality, health, fertility, and family planning are modeled
using Bayesian methods (Alkema and New 2014; Alkema et al. 2017;
Alexander and Alkema 2018; Cahill et al. 2018; Wakefield et al. 2019).

The majority of models used in these contexts are Bayesian hierarchical
frameworks, with a combination of systematic components capturing
theoretical relationships or relationships with key covariates, and a
temporal component which allows for trends over time to be smoothed and
projected forward. For example, Cahill et al. (2018) presents the Family
Planning Estimation Model (FPEM), a country-level model of contraceptive
use rates among married women of reproductive age from 1990-2020. The
model is based on a logistic curve, which assumes that adoption of
contraception is expected to start slowly, speed up, then slow down
before reaching an asymptote. Specific parameters of the curve are
data-driven and modeled hierarchically, with deviations from the
expected rate of change modeled with an AR(1) process.

Another example is the `Bmat' model, discussed in Alkema et al. (2017),
which describes the estimation method used by the United Nations
Maternal Mortality Estimation Interagency Group (MMEIG) to produce
country-level trends in the all-cause maternal mortality ratio (MMR).
The model consists of a hierarchical model which models the non-HIV/AIDS
MMR as a function of the general fertility rate, average number of
skilled attendants at birth, and gross domestic product. In addition,
country-time-specific deviations are modeled as an ARMA(1,1) process,
allowing for noisy observed trends to be smoothed and projected forward.
The Bmat model is particularly relevant to this work, as the estimates
of all-cause maternal mortality are used as the `envelope' for our
cause-specific estimates, meaning that we constrain the total number of
maternal deaths to be consistent with those produced by the MMEIG.

In general, Bayesian hierarchical models are particularly suited to
problems of estimating global health indicators for a set of multiple
populations with varying data quality and availability. For example, the
use of informative priors, based on substantive and theoretical
knowledge of the process being modeled, is useful to obtain estimates in
populations where the level of missing data are high. In addition,
hierarchical structures allow for information about mortality and other
health trends to be shared across similar populations that may have
varying amounts of data availability. Finally, Bayesian models aid in
the combination of multiple data sources with varying types of data
quality.

\hypertarget{estimation-of-cause-specific-mortality}{%
\subsection{Estimation of cause-specific
mortality}\label{estimation-of-cause-specific-mortality}}

In terms of efforts to model and estimate specific causes of death, and
with multiple causes being estimated in the same model (that is, the
estimation of cause of death distributions), the existing literature
mostly focuses on estimating cause-specific child mortality or maternal
mortality. This is probably partly driven by research agendas in the
Millennium Development (MDG) and Sustainable Development Goal (SDG)
eras, of which two important indicators are child and maternal health
(as part of SDG 3).

A range of different statistical approaches have been used to model
cause of death distributions. For example, Liu et al. (2016) uses a
multinomial logistic regression to model cause proportions of under-five
child mortality, while making various data adjustments to account for
varying data quality and underlying cause profiles. The resulting cause
proportions are then applied to the `envelope' (all-cause) estimates
produced by the methodology described in Alkema and New (2014) to get
death counts. Other approaches to modeling causes of death tend to model
causes separately and then rescale or perform post-hoc adjustments to
ensure the sum of the deaths by causes adds up to some predefined total.
In particular, published estimates of causes specific deaths produced as
part of the Global Burden or Disease Study (Naghavi et al. 2017; Wang et
al. 2017) rely on a multi-stage ensemble modeling technique which models
each cause of death in a similar but separate way. In contrast,
Schumacher et al. (2020) present a model that allows for cause- and
age-specific child mortality to be estimated from sample registration
data, with cause-specific and total deaths being estimated all within
the one framework.

Say et al. (2014) present estimates of maternal deaths by cause based on
a Bayesian hierarchical framework which models the underlying
proportions of death as a multivariate logistic Normal distribution.
Assumptions about data quality and coverage are based on the source of
the data, and HIV/AIDS maternal deaths are modeled separately. The model
we propose in this paper improves on Say et al. (2014) in several ways;
in particular, we explicitly account for the type of data,
down-weighting data sources of lower quality, and include a much larger
data set of observations of causes of maternal deaths, including those
from subnational sources, single-causes studies, and potentially
multiple data sources for the same country-year.

\hypertarget{goals-of-estimation}{%
\section{\texorpdfstring{Goals of estimation
\label{section_goals}}{Goals of estimation }}\label{goals-of-estimation}}

A maternal death is defined by the World Health Organization as `the
death of a woman while pregnant or within 42 days of termination of
pregnancy, irrespective of the duration and site of the pregnancy, from
any cause related to or aggravated by the pregnancy or its management
but not from accidental or incidental causes' (WHO 2020). We are
interested in obtaining estimates of the proportion of total maternal
deaths in a particular country and year by each of the 7 main cause
categories (as defined below), including additional breakdowns of 3 main
categories into sub-cause categories. Once estimates of the cause of
death distribution for each country-year are obtained, they are
aggregated across both geographic and time dimensions to obtain
estimates of cause of death distribution by SDG region for a specific
time period of interest (2009-2017).

\hypertarget{main-cause-of-death-categories}{%
\subsection{\texorpdfstring{Main cause of death categories
\label{subsection_causes}}{Main cause of death categories }}\label{main-cause-of-death-categories}}

There are many potential cause of death group classifications that we
could consider estimating. Motivated by previous work and priorities of
health policy agendas (Say et al. 2014), we estimate the proportion of
maternal deaths from each of the following seven cause categories:

\begin{itemize}
\tightlist
\item
  Abortion (ABO)
\item
  Embolism (EMB)
\item
  Hypertension (HYP)
\item
  Hemorrhage (HEM)
\item
  Sepsis (SEP)
\item
  Other direct causes (DIR)
\item
  Other indirect causes (IND)
\end{itemize}

Of particular note are maternal deaths related to HIV/AIDS, which are
encompassed in the indirect causes category. However, due to the
substantially different epidemiological profile of this death category,
as well as vastly different trends over time, we follow previous work on
mortality estimation (Alkema and New 2014; Say et al. 2014; Alkema et
al. 2017) in restricting our goal to be estimating the proportion of
non-HIV/AIDS maternal deaths by cause. Estimates of HIV/AIDS maternal
deaths produced elsewhere (UNAIDS 2017; WHO et al. 2019) are then
incorporated into the final cause of death distributions (as part of the
indirect cause group). See Section \ref{section_hiv} for more details.

\hypertarget{cause-of-death-subcategories}{%
\subsection{\texorpdfstring{Cause of death subcategories
\label{subsection_subcauses}}{Cause of death subcategories }}\label{cause-of-death-subcategories}}

In addition to the 7 main cause of death categories listed above, we are
interested in estimating further sub-cause breakdowns of hemorrhage,
sepsis and other direct cause groups. For hemorrhage (HEM) and sepsis
(SEP) deaths, we estimate the following breakdowns based on the timing
of death:

\begin{itemize}
\tightlist
\item
  Ante-partum (ANT)
\item
  Intra-partum (INT)
\item
  Post-partum (POS)
\end{itemize}

For other direct maternal deaths (DIR), we further break down the
category into four sub-categories:

\begin{itemize}
\tightlist
\item
  Anesthesia (ANE)
\item
  Obstructed labour (OBS)
\item
  Obstetric trauma (OBT)
\item
  Other causes (OTH)
\end{itemize}

\begin{figure}
\scalebox{0.8}{
\linespread{1}
\begin{tikzpicture}

% top part
\node (mat) at (9, 6) [rectangle, draw = black, fill = white] {maternal deaths};
\node (nonaids) at (9, 3.5) [rectangle, draw = black, fill = white] {non-HIV/AIDS};
\node (aids) at (18, 3.5) [rectangle, draw = black, fill = white] {HIV/AIDS};
\draw[thick, ->] (mat.south) -- (nonaids.north);
\draw[thick, ->] (mat.south) -- (aids.north);
\draw[red, thick, ->] (nonaids.south) -- (9, .5);
\node(nfi1) at (3, 4.5) [rectangle, dashed, draw = black, fill = none, align = center] { no further \\information};
\draw[dashed, ->] (mat.south) -- (nfi1.north);
\node(nfi2) at (5, 1.5) [rectangle, dashed, draw = black, fill = none, align = center] {no further \\information};
\draw[dashed, ->] (nonaids.south) -- (nfi2.north);

% main causes
\node at (1.5, 0.8) [text = red] {main category distribution};
\fill [red!30] (-1, -.5) rectangle (19, .5); 
\node (ABO) at (0, 0) [rectangle, draw=black, fill=white] {ABO};
\node (DIR) at (3, 0) [rectangle, draw=black, fill=white] {DIR};
\node (EMB) at (6, 0) [rectangle, draw=black, fill=white] {EMB};
\node (HEM) at (9, 0) [rectangle, draw=black, fill=white] {HEM};
\node (HYP) at (12, 0) [rectangle, draw=black, fill=white] {HYP};
\node (SEP) at (15, 0) [rectangle, draw=black, fill=white] {SEP};
\node (IND) at (18, 0) [rectangle, draw=black, fill=white] {IND};
\draw[thick, ->] (aids.south) -- (IND.north);

% DIR subcauses
\node at (3.5, -5.5) [text = orange] {DIR subdistribution};
\fill [orange!30] (2.6, -1) rectangle (4.4, -5);
\node (DIRobs) at (3.5, -1.5) [rectangle, draw=black, fill = white] {DIR\textsubscript{obs}};
\node (DIRane) at (3.5, -2.5) [rectangle, draw=black, fill = white] {DIR\textsubscript{ane}};
\node (DIRobt) at (3.5, -3.5) [rectangle, draw=black, fill = white] {DIR\textsubscript{obt}};
\node (DIRoth) at (3.5, -4.5) [rectangle, draw=black, fill = white] {DIR\textsubscript{oth}};
\draw[orange, thick, ->] (DIR.south) -- (3.5, -1);
\node(nfi3) at (1, -2) [rectangle, dashed, draw = black, fill = none, align = center] {no further \\information};
\draw[dashed, ->] (DIR.south) -- (nfi3.north);

% HEM subcauses
\node at (9.5, -4.5) [text = OliveGreen] {HEM subdistribution};
\fill [OliveGreen!30] (8.5, -1) rectangle (10.5, -4);
\node (HEMante) at (9.5, -1.5) [rectangle, draw=black, fill = white] {HEM\textsubscript{ante}};
\node (HEMintra) at (9.5, -2.5) [rectangle, draw=black, fill = white] {HEM\textsubscript{intra}};
\node (HEMpost) at (9.5, -3.5) [rectangle, draw=black, fill = white] {HEM\textsubscript{post}};
\draw[OliveGreen, thick, ->](HEM.south) -- (9.5, -1);
\node(nfi4) at (7, -2) [rectangle, dashed, draw = black, fill = none, align = center] {no further \\information};
\draw[dashed, ->] (HEM.south) -- (nfi4.north);

% SEP subcauses
\node at (15.5, -4.5) [text = blue] {SEP subdistribution};
\fill [blue!30] (14.5, -1) rectangle (16.5, -4);
\node (SEP_ante) at (15.5, -1.5) [rectangle, draw = black, fill = white] {SEP\textsubscript{ante}};
\node (SEP_intra) at (15.5, -2.5) [rectangle, draw = black, fill = white] {SEP\textsubscript{intra}};
\node (SEP_post) at (15.5, -3.5) [rectangle, draw = black, fill = white] {SEP\textsubscript{post}};
\draw[blue, thick, ->] (SEP.south) -- (15.5, -1);
\node(nfi5) at (13, -2) [rectangle, dashed, draw = black, fill = none, align = center] {no further \\information};
\draw[dashed, ->] (SEP.south) -- (nfi5.north);

\end{tikzpicture}
}

\caption{Organization of maternal deaths into analysis categories. Non-HIV/AIDS deaths are classified into 7 main cause of death categories, of which hemorrhage (HEM), sepsis (SEP),  and other direct causes (DIR) are classified further. After estimation of the main distribution using only non-HIV/AIDS observations, externally obtained estimates of HIV/AIDS are added to the other indirect causes (IND) category to give the complete distribution.}
\label{fig:data_diagram}

\end{figure}
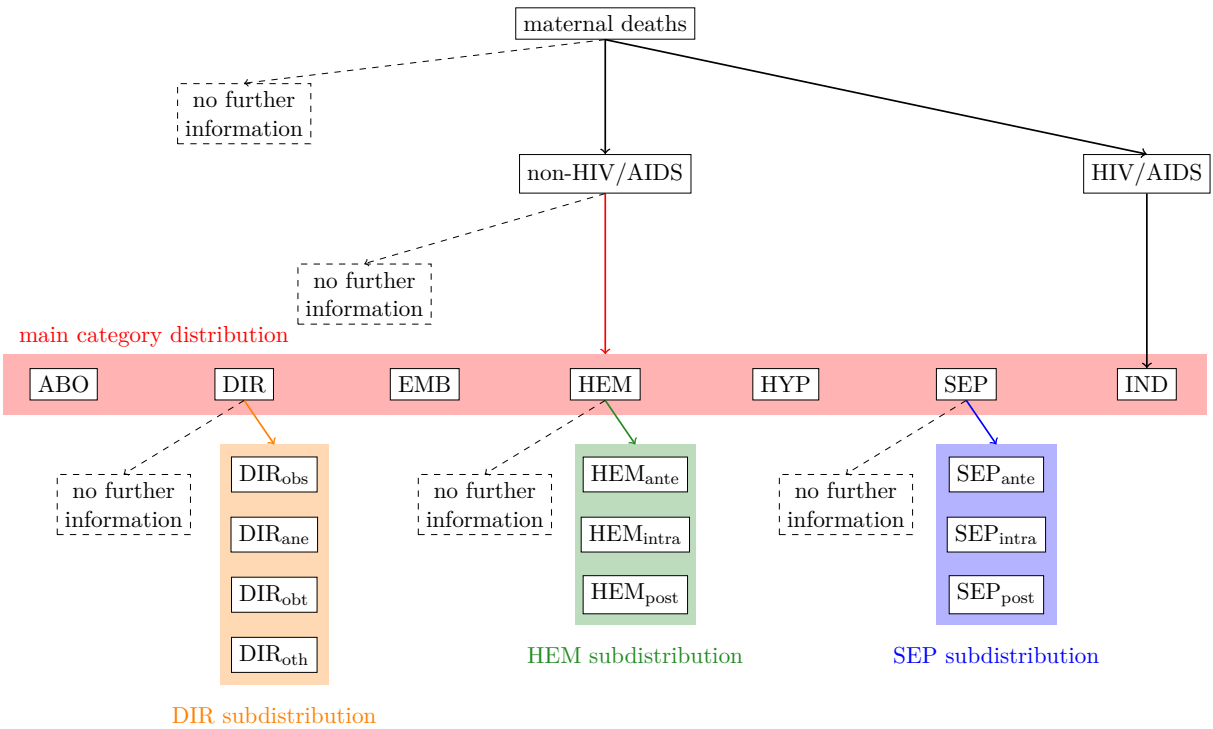

Thus, a total of 14 separate causes of death categories are estimated.
Figure \ref{fig:data_diagram} summarizes the classification of data
available on maternal deaths, and the cause of death categories to be
estimated. Maternal deaths are first classified as either deaths due to
HIV/AIDS or non-HIV/AIDS. Of the non-HIV/AIDS deaths, these can be
further classified into one of the 7 main cause of death categories. For
deaths due to hemorrhage (HEM), sepsis (SEP), and other direct causes
(DIR), we further classify into one of 10 sub-cause categories. In each
stage of the classification process, if no further information is
available, these deaths are excluded from the analysis.

\hypertarget{data}{%
\section{\texorpdfstring{Data \label{section_data}}{Data }}\label{data}}

\hypertarget{overview-of-data-sources}{%
\subsection{Overview of data sources}\label{overview-of-data-sources}}

Data on maternal deaths by cause come from three main sources: Civil
Registration and Vital Statistics (CRVS) systems; `grey literature,'
which refers to government reports, technical reports and other
non-peer-reviewed publications; and `studies,' which were the result of
a large-scale systematic literature review. CRVS systems and grey
literature provide national-level data. Data from studies, on the other
hand, can be from a number of different geographic levels. While some
may be at the national level, many refer to subnational areas, for
example Administrative 1 level (ADM1, state/province), Administrative 2
level (county), or lower. Indeed, in many cases, studies may only report
observation from a single hospital or group of hospitals or other health
institutions.

\hypertarget{overview-of-data-availability}{%
\subsection{Overview of data
availability}\label{overview-of-data-availability}}

\begin{table}[!h]

\caption{\label{tab:data_table_1}Breakdown of data availability by SDG region and source.}
\centering
\resizebox{\linewidth}{!}{
\begin{tabular}[t]{>{\raggedright\arraybackslash}p{4cm}>{\raggedleft\arraybackslash}p{2cm}>{\raggedleft\arraybackslash}p{2cm}>{\raggedleft\arraybackslash}p{2cm}>{\raggedleft\arraybackslash}p{2cm}>{\raggedleft\arraybackslash}p{2cm}>{\raggedleft\arraybackslash}p{2cm}>{\raggedleft\arraybackslash}p{2cm}}
\toprule
\multicolumn{3}{c}{ } & \multicolumn{4}{c}{Number of observed country-years by data sources} & \multicolumn{1}{c}{ } \\
\cmidrule(l{3pt}r{3pt}){4-7}
SDG  Region & Number of countries & Number of observed country-years & CRVS & Grey Literature & Studies ADM1 or higher & Studies Below ADM1 & Prop. country-years missing 2+ causes\\
\midrule
\cellcolor{gray!6}{Central and Southern Asia} & \cellcolor{gray!6}{12} & \cellcolor{gray!6}{122} & \cellcolor{gray!6}{31} & \cellcolor{gray!6}{16} & \cellcolor{gray!6}{2} & \cellcolor{gray!6}{73} & \cellcolor{gray!6}{0.533}\\
Europe and Northern America & 36 & 260 & 252 & 8 & 0 & 0 & 0.738\\
\cellcolor{gray!6}{Northern Africa and Western Asia} & \cellcolor{gray!6}{18} & \cellcolor{gray!6}{91} & \cellcolor{gray!6}{81} & \cellcolor{gray!6}{6} & \cellcolor{gray!6}{0} & \cellcolor{gray!6}{4} & \cellcolor{gray!6}{0.538}\\
Oceania excl. Australia and New Zealand & 3 & 7 & 4 & 1 & 0 & 2 & 0.714\\
\cellcolor{gray!6}{Sub-Saharan Africa} & \cellcolor{gray!6}{28} & \cellcolor{gray!6}{151} & \cellcolor{gray!6}{18} & \cellcolor{gray!6}{20} & \cellcolor{gray!6}{15} & \cellcolor{gray!6}{98} & \cellcolor{gray!6}{0.642}\\
\addlinespace
Latin America and the Caribbean & 31 & 218 & 207 & 10 & 0 & 1 & 0.339\\
\cellcolor{gray!6}{Australia and New Zealand} & \cellcolor{gray!6}{2} & \cellcolor{gray!6}{26} & \cellcolor{gray!6}{16} & \cellcolor{gray!6}{10} & \cellcolor{gray!6}{0} & \cellcolor{gray!6}{0} & \cellcolor{gray!6}{0.846}\\
Eastern and South-Eastern Asia & 12 & 75 & 46 & 9 & 2 & 18 & 0.467\\
\bottomrule
\end{tabular}}
\end{table}

Over the period of interest (2009--2017), at least one observation of
cause-specific maternal mortality were available for 142 of the 183 UN
member countries. Table \ref{tab:data_table_1} provides a breakdown of
the available data by SDG region. The majority of available country-year
observations, roughly 70\%, come from CRVS data. Most notably these are
observations from Europe and Northern America or Latin America and the
Caribbean. In other regions, data from the grey literature and studies
play a particularly important role. For instance, as Table
\ref{tab:data_table_1} demonstrates, the vast majority of available data
for Sub-Saharan Africa and Central and Southern Asia come from various
subnational studies.

It is worth noting that it is not uncommon for a country-year to report
maternal mortality counts for only a subset of the cause categories that
we are interested in estimating. As illustrated in the rightmost column
of Table \ref{tab:data_table_1}, the majority of the country-year
observations in each region have two or more causes missing.

\hypertarget{classification-into-cause-of-death-groups-and-subgroups}{%
\subsection{Classification into cause of death groups and
subgroups}\label{classification-into-cause-of-death-groups-and-subgroups}}

Where possible, deaths are assigned into one of the 14 cause categories
listed in Section \ref{subsection_causes} and \ref{subsection_subcauses}
above. This is straightforward when deaths are classified according to
ICD-10 codes, which is the case for the majority of data (encompassing
all CRVS and some studies and grey literature data). The assignment of
ICD-10 codes to each of the 14 cause categories is summarized in
Appendix \ref{appendix_icd10}.

However, for the studies and grey literature data, classification into
cause categories is more complex. In many studies the cause of death is
referred to using a free text description, rather than an ICD-10 code.
Sometimes the free text can be clearly translated into an ICD-10 code or
be identified as one of the broader level ICD-10 cause groups defined by
the WHO (WHO 2012) and, hence, designated into one of the seven main
cause categories. Other times, the free text describes more than a
single ICD-10 code or cause group associated with a number of maternal
deaths. In such cases we rely on the clinician's expert interpretation
of the correct proportion of maternal deaths corresponding to each code
or cause group. Cases where the cause of death could not be assigned
(for example, instances where only the likely organ system causing the
death could be identified from free text) were excluded from the
analysis.

\hypertarget{other-data-sources}{%
\subsection{Other data sources}\label{other-data-sources}}

In addition to data on causes of maternal death, we make use of a number
of other data sources in the model. We use annual all-cause female
mortality estimates from the UN World Population Prospects (WPP) 2019
edition to assess the coverage, and subsequently the `usability,' of
death data observed by country (UNPD 2019). We also employ the UN
Maternal Mortality Estimation Inter-Agency Group's (MMEIG) 2019
estimates on maternal mortality to obtain an estimate for the total
number of maternal deaths, and the number of maternal deaths due to
HIV/AIDS (WHO et al. 2019).

Additional information on data adjustments related to the treatment of
zero reported deaths, HIV/AIDS deaths, and multi-year studies are
detailed in Appendix \ref{appendix_data_adjustments}.

\hypertarget{model}{%
\section{\texorpdfstring{Model
\label{section_model}}{Model }}\label{model}}

\hypertarget{overview}{%
\subsection{Overview}\label{overview}}

We begin with grouping the observed non-HIV/AIDS cause-specific death
counts for each country into 7 main cause of death categories, and we
model these as coming from a multinomial distribution with 7 categories.
The relative proportions in the multinomial distribution then depend
primarily on regional and country-level differences. Additionally, we
include an error term applied only to lower quality data to allow for
deviation from the assumed true country mean to lessen the impact of
potentially misrepresentative observations.

As detailed in Section \ref{section_goals}, subdistributions for certain
categories are also calculated. The DIR, HEM, and SEP counts are further
divided into separate multinomial distributions representing more
granular classifications of cause of death. Similar to the main cause
distribution, these are modeled as the sum of regional and country-level
differences.

The true main cause of death distribution and the subcause distributions
for each country are calculated using the estimated regional and a
weighting of the estimated country-level differences. In order to
reflect increased uncertainty in countries where data coverage is low,
the weight given to the country's specific estimated term depends on the
coverage of the available data in that country.

\hypertarget{notation}{%
\subsection{Notation}\label{notation}}

For \(i = 1, \dots, N\), let
\(\mathbf{y_i} = (y_{i,1}, \dots, y_{i,7})\) be the observed
non-HIV/AIDS maternal deaths by cause, where \(y_{i,j}\) is the number
of deaths in the \(i\)th observation attributed to cause group \(j\),
for \(j \in \{\text{ABO, EMB, HEM, SEP, DIR, IND, HYP}\}.\) Define the
total number of observed non-HIV/AIDS maternal deaths for the \(i\)th
observation to be \(d_i\).

Point estimates from the MMEIG of the HIV/AIDS-omitted total number of
maternal deaths in a country-year \(ct\) are denoted as
\(\mathring{d}_{ct}\).

\hypertarget{model-for-observed-cause-proportions}{%
\subsection{Model for observed cause
proportions}\label{model-for-observed-cause-proportions}}

Deaths in each observation are modeled using a multinomial distribution.
The log-ratios of proportions are taken to be the sum of an intercept,
region effect, and country effect. However, for observations from CRVS
or studies data, an additional adjustment is applied to allow for
deviation from the true country mean. This adjustment allows us to
partially mitigate the impact of potentially misrepresentative
observations that arise from lower quality data.

For observation \(i\), the multinomial proportions
\((p_{i,1}, \dots, p_{i, 7})\) are modeled as \begin{equation*}
\begin{gathered}
\mathbf{y_i} \sim \text{Multinomial}(d_i, \mathbf{p_i}) \\
\mathbf{p_i} = (p_{i,1}, \dots, p_{i, 7}) \\
\log \left(\frac{p_{i,j}}{p_{i,7}}\right) = \beta_{0,j} +  \beta_{r(c(i)), j} + u_{c(i),j} + q_{i,j}, \\
\end{gathered}
\end{equation*} where \(c(i)\) refers to country of observation \(i\),
and \(r(c(i))\) corresponds to the region of that country. The log-ratio
of proportions for category \(j\) relative to category 7 (HYP) is taken
to be the sum of an intercept term \(\beta_{0, j}\), a region effect
\(\beta_{r(c(i)), j}\), a country effect \(u_{c(i),j}\) and a data
quality adjustment term \(q_{i,j}\).

The region effects are pooled with a global variance term. For all
regions \(r = 1, \dots, R\), \begin{equation*}
\begin{gathered}
\beta_{r,j} \sim \text{Normal} (0, \sigma_{\beta}^2) \\
\sigma_{\beta} \sim \text{Normal} (0, 1^2).
\end{gathered}
\end{equation*} The regions \(r = 1, \dots, R\) are the same as those
used by Say et al. (2014), which broadly aim to group countries that are
believed to be epidemiologically similar. This region classification is
distinct from the SDG region system used for aggregating and reporting
results. More details on the modeling region classification are given in
Appendix \ref{appendix_regions}.

\hypertarget{multinomial-likelihood-and-missing-values}{%
\subsubsection{Multinomial likelihood and missing
values}\label{multinomial-likelihood-and-missing-values}}

In an ideal situation where counts for all categories are recorded, each
\(\mathbf{y_i}\) is treated as a 7-category multinomial observation with
probabilities \((p_{i, 1}, \dots, p_{i, 7})\). For observation \(i\),
let the ratio of proportion \(j\) to the reference category be denoted
\(g_{ij}\). That is, \[
\frac{p_{i,j}}{p_{i,7}} = g_{i,j} = \exp(\beta_{0,j} +  \beta_{r(c(i)), j} + u_{c(i),j} + q_{i,j}).
\] The probabilities can then be expressed \[
p_{i,j} = \frac{g_{i,j}}{\sum_{k=1}^7 g_{i,k}},
\] and the corresponding multinomial likelihood \(L_M\)for all \(N\)
observations is \begin{equation}
\label{eq:full-multinomial}
L_M =  \prod_{i=1}^N \prod_{j=1}^7 p_{i,j}^{y_{i,j}} = \prod_{i=1}^N \prod_{j=1}^7 \left(\frac{g_{i,j}} {\sum_{k=1}^7 g_{i,k} } \right)^{y_{ij}}.
\end{equation}

However, as stated in Section \ref{section_data}, there are observations
for which certain categories' counts are considered to be missing. In
such cases where we believe an apparent zero count is some category
\(k\) is unreliable, we wish to treat \(y_{i, k}\) as unknown instead.

We can accomplish this by treating the observation as a multinomial
observation with a reduced number of categories. For example, if \(j=1\)
(ABO) is missing for observation \(i\), then the likelihood contribution
for that observation would instead be \[
\prod_{j=2}^7 \grave{p}_{i,j}^{y_{i,j}} = \prod_{j=2}^7 \left(\frac{g_{i,j}} {\sum_{k=2}^7 g_{i,k} } \right)^{y_{ij}}.
\] where the probabilities \((\grave{p}_{2,j}, \dots, \grave{p}_{i,7})\)
are the original probabilities rescaled to sum to 1.

An appropriately reduced multinomial is used for every observation with
any combination of missing categories.

\hypertarget{country-specific-effect-and-correlations-across-causes}{%
\subsubsection{Country-specific effect and correlations across
causes}\label{country-specific-effect-and-correlations-across-causes}}

We allow for the possibility of correlations between death categories in
countries' cause of death distributions. Intuitively, because of
co-morbidities and common, systematic underlying factors affecting
maternal health, we would expect certain death categories to co-occur.

In order to capture correlations in countries' cause of death
distributions, the country effects \(u_{c,j}\) are modeled as
multivariate normal with a common 6x6 covariance matrix \(\Sigma\). We
decompose \(\Sigma\) into its correlation matrix \(\Omega\) and diagonal
matrices of variance terms,
\(\Sigma = \text{diag}(\mathbf{v})\, \Omega \, \text{diag} (\mathbf{v})\).
An \(\text{LKJ}(1)\) prior is used for \(\Omega\) (Lewandowski,
Kurowicka, and Joe 2009; Stan Development Team 2019) . For each country
\(c = 1, \dots, C\), \begin{equation*}
\begin{gathered}
(u_{c,1}, \dots, u_{c, 6}) \sim \text{MVN} (\mathbf{0}, \Sigma) \\
\Sigma = \text{diag}(\mathbf{v}) \Omega \text{diag} (\mathbf{v}) \\
\Omega \sim \text{LKJ}(1) \\
\mathbf{v} \sim \text{Normal} (0, 3).
\end{gathered}
\end{equation*}

\hypertarget{data-quality-adjustment}{%
\subsubsection{\texorpdfstring{Data-quality adjustment
\label{subsection_model_dataquality}}{Data-quality adjustment }}\label{data-quality-adjustment}}

Even where cause of death data are available, the data may be
misrepresentative due to systematic deficiencies and biases in data
collection processes. In our model, we classify observations into 4 data
quality `types', and incorporate this information into the model by
including an extra error term for observations from lower quality types,
partially mitigating the impact of these observations on the country
means.

The data quality type classification of each observation depends on the
source and estimated coverage of the observation. Type 1 observations
consist only of observations arising from grey literature. These are
considered to be of the highest quality, and no additional adjustment is
applied to these data.

Observations collected from studies are assigned to be Type 4, the
lowest quality, since the included studies may from subnational regions,
and also may not aim to be a comprehensive surveying of all causes of
maternal death.

Observations from CRVS data can be classified as Type 2, 3, or 4,
depending on the estimated coverage of the data. We follow a similar
approach to Say et al. (2014) in defining a `usability' index, which is
a function of the presence of ill-defined deaths, the coverage of the
observed number of deaths, and the presence of contributory-cause
misclassification.

For observations where the number of observed maternal deaths is no more
than 5, the usability index is calculated using the proportion of
ill-defined deaths denoted \(p_{i}^{\text{ill}}\), and the all-cause
(maternal and otherwise) female death coverage \(C_i\) of the
observation. This is calculated as the ratio of all-cause death count in
the CRVS system to the all-cause death count estimate from WPP 2019
(UNPD 2019). In particular, the usability for CRVS observation \(i\),
\(\nu_i\) is calculated as \[
\nu_i = \frac{d_i}{\mathring{d}_{ct(i)}} (1 - p_{i}^{\text{ill}}).
\] If there are more than 5 maternal deaths in an observation, then we
also consider the proportion \(p_{i}^{\text{contr}}\) of maternal deaths
attributed to contributory causes. Maternal deaths should always be
classified with a main ICD-10 code that is an underlying cause of death,
not a contributory cause of death. If contributory causes are listed as
the main cause of death, then this suggests CRVS systems of a lower
quality. In particular, where there are more than 5 maternal deaths
observed, \(\nu_i\) is calculated as \[
\nu_i = \frac{d_i}{\mathring{d}_{ct(i)}} (1 - p_{i}^{\text{ill}}) (1 -  p_{i}^{\text{contr}})
\] The usability index \(\nu_i\) is then used to classify CRVS data into
types 2, 3, or 4 as follows:

\begin{itemize}
\tightlist
\item
  the observation is classified as type 2 if \(\nu_i > 85\%\) and is one
  of three consecutive years with \(\nu_i > 60 \%\),
\item
  type 3 if \(65\% < u_i \leq 85\%\) and is one of three consecutive
  years with \(u_i > 60 \%\), and
\item
  type 4 otherwise.
\end{itemize}

\begin{table}

\caption{\label{tab:unnamed-chunk-2}Summary of data quality type classification. Highest quality data is classified as type 1, and lowest quality data is classified as type 4. Data quality is assessed based on data source and a calculated usability index.}
\centering
\fontsize{10}{12}\selectfont
\begin{tabular}[t]{rl}
\toprule
Type & Sources\\
\midrule
\cellcolor{gray!6}{1} & \cellcolor{gray!6}{most grey literature observations}\\
2 & some grey lterature observations; high usability CRVS\\
\cellcolor{gray!6}{3} & \cellcolor{gray!6}{medium usability CRVS}\\
4 & low usability CRVS; studies\\
\bottomrule
\end{tabular}
\end{table}

For observations arising from type 2, 3, or 4 data sources, a data
adjustment term \(q_{i,j}\) is individually realized from a Normal
distribution for each observation, with an estimated variance term
dependent on the type of data source. For each observation \(i\) and
cause \(j\), \begin{equation*}
\begin{gathered}
q_{i, j} 
\begin{cases}
  = 0 & \text{if type}(i) = 1 \\
  \sim \text{Normal} (0, \sigma^2_{\text{type}(i)}) & \text{if type}(i) = 2, 3, 4
\end{cases} \\
\sigma_{\text{type}} \sim \text{Normal} (0, 0.25^2).
\end{gathered}
\end{equation*}

\hypertarget{estimation-of-true-proportions}{%
\subsection{Estimation of true
proportions}\label{estimation-of-true-proportions}}

Although some observations may have large death counts, they may only
represent a fraction of the true number of maternal deaths in that
country-year, which we take to be the MMEIG point estimate
\(\mathring{d}_{ct}\). In these cases where a large fraction of the
deaths are unrecorded or unclassified, we may have an unduly high degree
of certainty in our estimate of \(p_{i,j}\). We therefore apply a
weighting scheme to better reflect our uncertainty where data coverage
is low. For country \(c\), let \(w(c) \in [0,1]\) be the maximum
coverage among its observations. If \(ct(i)\) refers to the country-year
of observation \(i\), then \[
w(c) := \max_{i:\,c(i) = c} \frac{d_i}{\mathring{d}_{ct(i)}}.
\] The weight \(w(c)\) is then used to regulate the contribution of the
estimated country effect \(u_{c,j}\). For countries with only
low-coverage observations (\(w(c)\) is low), the country-specific
information is considered incomplete, and the estimate should therefore
lean more heavily towards the region mean rather than the observed
country mean. The estimate of the true HIV/AIDS-omitted cause of death
distribution is calculated as \begin{equation*}
\begin{gathered}
\log \left(\frac{p^\ast_{c,j}}{p^\ast_{c,7}}\right) = \beta_{0,j} + \beta_{r(c), j} + w_{c} \cdot u_{c,j} + (1-w_{c}) \cdot \tilde{u}_{c, j} \\
\tilde{\mathbf{u}}_{c} \overset{\text{RNG}}{\sim} \text{MVN} (0, \Sigma),
\end{gathered}
\end{equation*} where \(\tilde{\mathbf{u}}_c\) is a new generic
realization of the country effect.

\hypertarget{incorporating-hivaids-deaths}{%
\subsection{\texorpdfstring{Incorporating HIV/AIDS deaths
\label{section_hiv}}{Incorporating HIV/AIDS deaths }}\label{incorporating-hivaids-deaths}}

Deaths from HIV/AIDS are treated separately. Recall that
\(\mathring{d}_{ct}\) denotes the HIV/AIDS-omitted MMEIG estimate of
maternal deaths in country-year \(ct\). Let
\(\mathring{d}_{ct, \text{HIV}}\) denote the MMEIG point estimate of the
number of HIV/AIDS deaths in country-year \(ct\), and let
\(\mathring{p}_{ct, \text{HIV}} = \mathring{d}_{ct, \text{HIV}} / ( \mathring{d}_{ct, \text{HIV}} + \mathring{d}_{ct})\)
denote the MMEIG-estimated proportion of maternal deaths attributed to
HIV/AIDS.

Similarly, let
\(\hat{p}_{i, \text{HIV}} = d_{i, \text{HIV}} / (d_i + d_{i, \text{HIV}})\),
where \(d_i\) and \(d_{i, \text{HIV}}\) denote the analogous quantities
from observation \(i\).

Define
\(\sigma_{\text{HIV}} = \text{sd}(\hat{p}_{i, \text{HIV}} - \mathring{p}_{ct(i), \text{HIV}})\).
Then \begin{equation*}
\begin{gathered}
p'_{ct, \text{HIV}} \overset{\text{RNG}}{\sim} \text{Normal} (\mathring{p}_{ct, \text{HIV}}, (\mathring{p}_{ct, \text{HIV}} \cdot \sigma_{\text{HIV}})^2) T(0, 1) \\ 
d'_{ct, \text{HIV}} = p'_{ct, \text{HIV}} \cdot (d_{ct, \text{HIV}} + \mathring{d}_{ct}). \\ 
\end{gathered}
\end{equation*}

We then add the HIV/AIDS deaths \(d'_{ct, HIV}\) to the to the IND
group, and recalculate proportions to obtain the final
HIV/AIDS-inclusive country-year distributions
\((p'_{ct, 1}, \dots, p'_{ct, 7})\). We do this by converting the
HIV/AIDS-omitted proportions \(p^\ast_{c,j}\) into counts
\(d^\ast_{ct,j}\), adding the estimated HIV/AIDS counts appropriately,
and rescaling these counts into proportions: \begin{equation*}
\begin{gathered}
d^\ast_{ct, j} = p^\ast_{c,j} \cdot \mathring{d}_{ct} \\ 
d'_{ct, j} = 
  \begin{cases}
    d^\ast_{ct, j} + d'_{i, \text{HIV}}  & \text{if } j=\text{IND}\\
    d^\ast_{ct, j} & \text{otherwise} \\
  \end{cases} \\
p'_{ct,j} = \frac{d'_{ct,j}}{\sum_{l=1}^7 d'_{ct,l}}.
\end{gathered}
\end{equation*}

\hypertarget{calculating-regional-and-global-cause-of-death-distributions}{%
\subsection{Calculating regional and global cause of death
distributions}\label{calculating-regional-and-global-cause-of-death-distributions}}

Regional cause of death distributions are obtained by aggregating
country death counts. For (SDG) regions \(h = 1, \dots, H\) (distinct
from regions \(r = 1, \dots, R\) used above in the model), the
countries' HIV/AIDS-inclusive counts \(d'_{ct, j}\) are aggregated
accordingly to obtain regional counts for each cause \[
d'_{h, j} = \sum_{h(c) = h} d'_{ct, j}.
\] The counts are then normalized to give proportions \[
p'_{h,j} = \frac{d'_{h, j}}{\sum_{l=1}^7 d'_{h, l}}.
\]

Similarly, global estimates are obtained by aggregating regional death
counts and normalizing \begin{equation*}
\begin{gathered}
d'_{\text{global}, j} = \sum_{h=1}^H d'_{h, j} \\
p'_{\text{global},j} = \frac{d'_{\text{global}, j}}{\sum_{l=1}^7 d'_{\text{global}, l}}
\end{gathered}
\end{equation*}

\hypertarget{subcause-distribution-estimation}{%
\subsection{Subcause distribution
estimation}\label{subcause-distribution-estimation}}

Within the cause groups HEM, SEP, and DIR, we are interested in a finer
classification of cause of death. For each of these three main
categories \(j\), let \(K_j\) denote the number of subcategories. These
categories can be further subdivided as follows.

\begin{enumerate}
\def\labelenumi{\arabic{enumi}.}
\tightlist
\item
  Within the hemorrhage (HEM) category, deaths can be classified as
  antepartum (HEM\textsubscript{ante}), intrapartum
  (HEM\textsubscript{intra}), and postpartum (HEM\textsubscript{post})
  hemorrhage. \(K_{\text{HEM}} = 3\).
\item
  The sepsis (SEP) category can similarly be divided into antepartum
  (SEP\textsubscript{ante}), intrapartum (SEP\textsubscript{intra}), and
  postpartum sepsis (SEP\textsubscript{post}). \(K_{\text{SEP}} = 3\)
\item
  The direct cause (DIR) category can be divided into
  (DIR\textsubscript{obs}), (DIR\textsubscript{ane}),
  (DIR\textsubscript{obt}), and other direct causes
  (DIR\textsubscript{oth}). \(K_{\text{DIR}} = 4\).
\end{enumerate}

The subcategory estimation procedure is similar to the main categories'
procedure, with some key differences. First, there is no data quality
error term (\(q\)) in the estimation of the log ratio of proportions.
Second, in the estimation of true proportions, we modify the weights
that balance the estimated country effect \(u\) and the new realization
of the country effect, \(\tilde{u}\). In the main category model, we use
the \(w_c\) and \(1-w_c\), where \(w_c\) is the defined as the maximum
coverage of any single country-year observation for country \(c\). This
was to reflect the uncertainty that arises from not observing up to a
proportion of \(1-w_c\) of the maternal deaths in a country-year. An
analogous statistic for the subcategory distributions is not readily
available, since we don't know specifically the coverage of deaths in
some category \(j\). For simplicity, we continue to use \(w_c\) for the
purpose of reflecting unobserved deaths in the HEM, SEP, and DIR
subdistributions. However, an additional complication is that more
granular information is unavailable for some observations. For instance,
some deaths may be attributed to hemorrhage but no further information
is given about the timing, and therefore do not hold useful information
about the subdistribution.

To account for this the subcategory model, we use the weight
\(z_{c, j} w_c\) and \(1 - z_{c, j} w_c\), where \(z_{c, j}\) is defined
as the maximum proportion, among observations for country \(c\), of
deaths in category \(j\) that have subcategory information available.

\hypertarget{implementation-and-computation}{%
\subsection{Implementation and
Computation}\label{implementation-and-computation}}

\hypertarget{equivalent-poisson-likelhood}{%
\subsubsection{Equivalent Poisson
Likelhood}\label{equivalent-poisson-likelhood}}

We implement the multinomial model using an equivalent Poisson
likelihood. The Poisson implementation speeds up computation, and easily
allows for arbitrary combinations of missing values. We rely on a result
presented by Ghosh, Zhang, and Mukherjee (2006) which states that the
multinomial likelihood in Equation (\ref{eq:full-multinomial}) is
equivalent to the Poisson likelihood \[
L_P = \prod_{i=1}^N \prod_{j = 1}^7 (g_{ij} \exp \phi_i )^{y_{ij}} \exp (- g_{ij} \exp \phi_i),
\] where \(\phi_i\) have independent improper priors \(p(\phi_i)\).

Note that if, for each observation \(i\), some arbitrary subset of
categories \(J_i \subset \{1, \dots, 7\}\) is missing, and the desired
reduced multinomial is across the categories
\(\{ 1, \dots, 7\} \backslash J_i\), then the likelihood reduces to \[
L_P = \prod_{i=1}^N \prod_{j \in \{1, \dots, 7\} \backslash J_i} (g_{ij} \exp \phi_i )^{y_{ij}} \exp (- g_{ij} \exp \phi_i).
\] in which the unused terms in \(J_i\) can simply be omitted from the
likelihood while the other terms remain unchanged.

In practical terms, this means we treat non-missing count \(y_{ij}\) as
coming from a Poisson distribution, \[
y_{ij} \sim \text{Poisson} (g_{ij} \cdot \exp\phi_i),
\] omitting missing counts from the likelihood evaluation. Importantly,
this means we do not have to rescale the probabilities to 1 for the
reduced multinomial likelihood, which would require recalculation of the
denominator \((\sum_{j \not \in K_i} g_{ij})\) for each observation.

\hypertarget{computation-in-stan}{%
\subsubsection{Computation in Stan}\label{computation-in-stan}}

Posterior samples were obtained using Hamiltonian Monte Carlo
implemented in Stan via the \texttt{cmdstanr} R package version 0.2.0,
with 4 parallel chains of 6000 warmup iterations and 4000 sampling
iterations (Stan Development Team 2019; Gabry and Češnovar 2020).
Standard checks for \(\hat{R}\) and effective sample size were
performed. This computation was enabled in part by resources provided by
Compute Ontario (\url{https://computeontario.ca/}) and ComputeCanada
(\url{https://www.computecanada.ca/}).

\hypertarget{results}{%
\section{\texorpdfstring{Results
\label{section_results}}{Results }}\label{results}}

In this section we illustrate some results of the estimation process at
the regional level and also for the three case-study countries. We also
present results of model validation exercises.

\hypertarget{results-by-region}{%
\subsection{Results by region}\label{results-by-region}}

\begin{figure}
\centering
\includegraphics{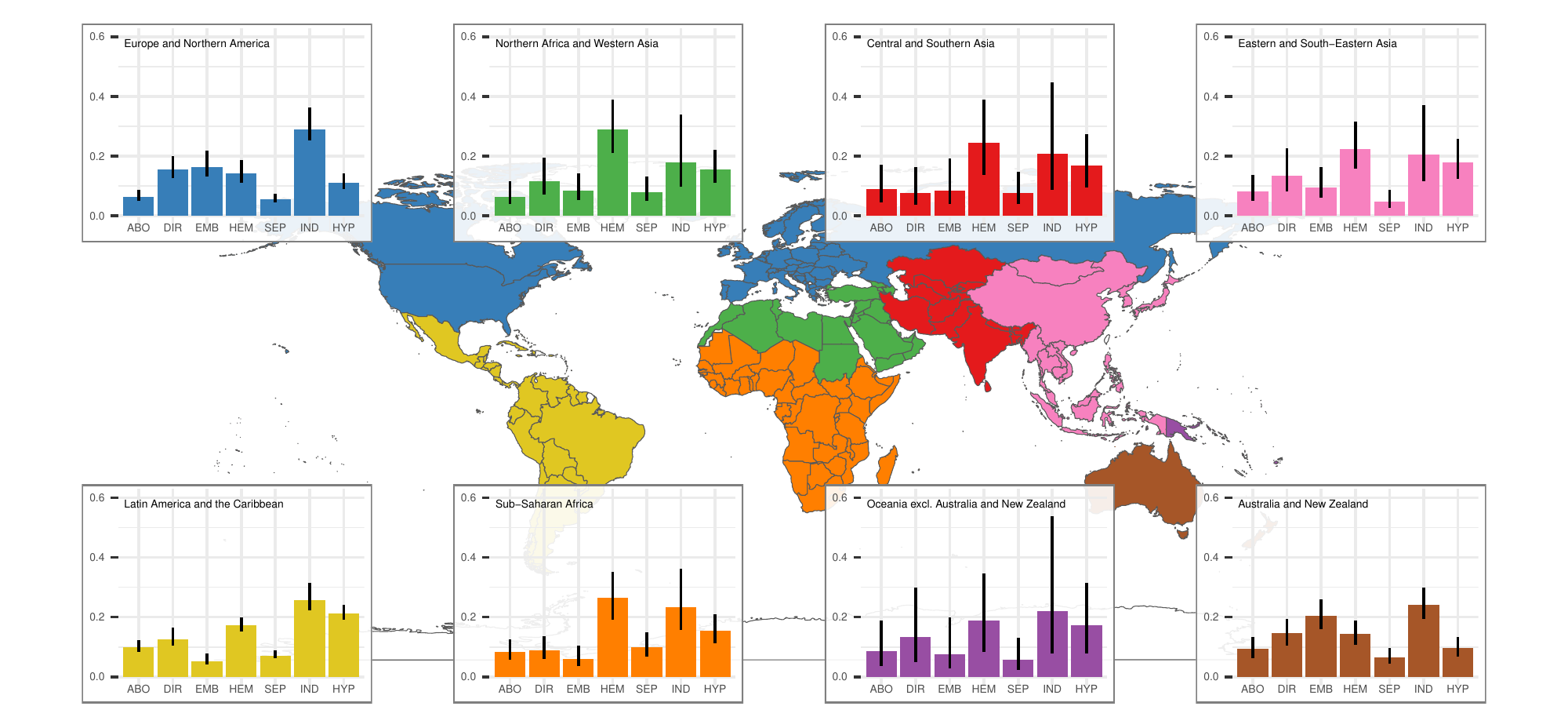}
\caption{\label{fig:fig_region}Estimates and 95\% credible intervals for
the maternal cause of death distributions for SDG regions}
\end{figure}

Figure \ref{fig:fig_region} illustrates the estimates and 95\% credible
intervals for the maternal cause of death distributions (showing the
seven main causes of death categories) for the seven SDG regions. The
results show substantial differences in the both the cause of death
distributions by region and also the uncertainty around the resulting
estimates.

The estimates for the Europe and North America, and Australia and New
Zealand regions, which both encapsulate high income countries, are
fairly similar. In particular, the proportion of deaths due to indirect
causes is relatively high, which is expected given these countries are
in the late stage of the epidemiological transition (Nair,
Nelson-Piercy, and Knight 2017). Additionally, the uncertainty around
the estimates in these regions is relatively low, which is a consequence
of the large amount of data available in these countries. The Latin
American and the Caribbean region also has a relatively high proportion
of indirect deaths and low uncertainty, but additionally has a high
proportion of deaths due to hypertension.

In other regions, such as in Africa and Asia, the proportion of deaths
due to hemorrhage is much higher, which partly reflects lack of access
to high quality health care and skilled attendants at birth (Prata et
al. 2011; Montgomery et al. 2014; Maduka and Ogu 2020). Larger
uncertainty intervals are mostly due to a lack of data available, apart
from in Oceania where the uncertainty is mostly driven by small
population sizes.

\hypertarget{case-studies}{%
\subsection{Case studies}\label{case-studies}}

\begin{figure}
\centering
\includegraphics{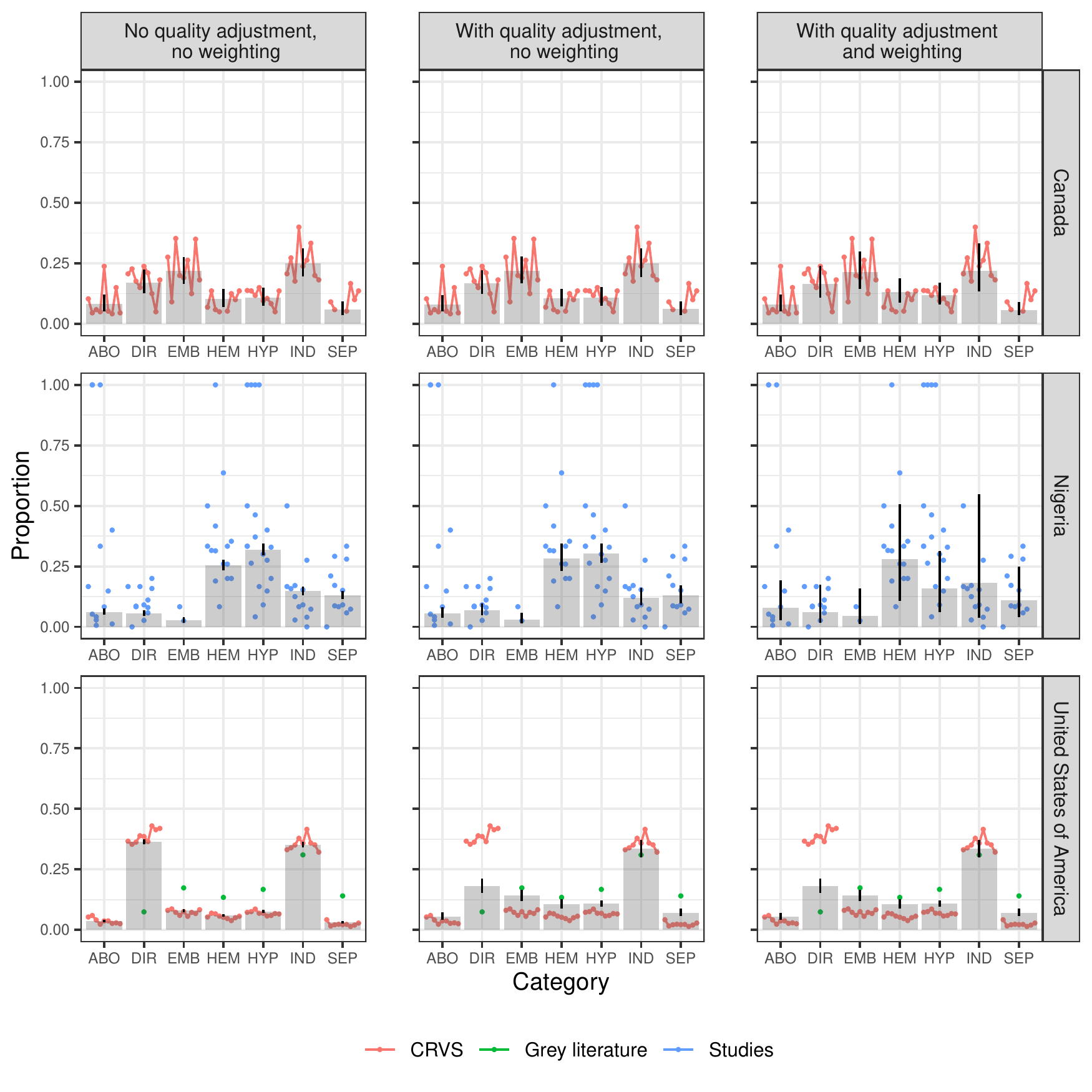}
\caption{\label{fig:fig_case_studies}Observed proportions (shown as
points), and estimates (shown as bars) with 95\% credible intervals for
three case study countries, using three different model set-ups. The
final model, shown in the right-most column, accounts for data coverage
and data quality.}
\end{figure}

Figure \ref{fig:fig_case_studies} illustrates the observed and estimated
cause of death distributions for the three case-study countries under
three different model set-ups, to illustrate the effect of different
model components. Country A is a country with relatively low maternal
mortality levels, with good-quality CRVS data available. Country B is a
country with relatively high maternal mortality but only has data
available from subnational studies. Country C has CRVS data with some
issues, in particular there are a high number of mis-classified `other
direct' maternal deaths. In addition, Country C also has one
high-quality data observation available from a maternal mortality
surveillance study.

In each of the graphs, the dots represent the observed proportions by
cause sequentially over 2009--2017. If the particular data source of
interest contained sequential years then the dots are joined together
with a line (for example, CRVS data in countries A and C). Note that the
multiple proportions equal to one observed in country B refer to
single-case studies, i.e.~studies where only one cause of death was
reported.

The columns of the figure show the implied estimates for three different
modeling set-ups. The first, shown on in the left-hand column shows the
results from a model with no quality adjustment and no weighting to
account for under-coverage. The middle column shows the results based on
a model with a quality adjustment but no weighting. And finally, the
right-hand column shows the results from a model with both the quality
and coverage adjustments, which is equivalent to the model described
above.

There are several observations to note. Firstly, the estimates for
Country A do not change substantially across the three model
alternatives. This reflects the fact that Country A's CRVS system is
considered of good quality and high coverage.

Secondly, looking at Country B, the addition of the quality term (from
the left to middle column) increases the uncertainty around the
estimates only slightly, after accounting for the study data being of
relatively poor quality. However, the addition of the weighting (from
the middle to right column) changes both the point estimates and
uncertainty intervals substantially. This is a consequence of the severe
under-coverage of the available data in Country B. Although there is one
relatively large study available in Country B with approximately 1400
deaths reported, this is still substantially lower than the estimated
66000 total number of maternal deaths here in 2015. As a consequence,
the final estimates for this country are weighted more heavily towards
its regional distribution (possibly away from its own observed
proportions), and the uncertainty around the estimates is much larger.

Finally, Country C is an example where the CRVS data have known issues,
but available data is supplemented with high-quality surveillance data
(`grey literature'). Going from left to right, the inclusion of a data
quality term has a substantial effect on the point estimates, as the
estimates are pulled to be more in line with the grey literature (in
particular for other direct deaths). The final step of weighting has
minimal impact as the coverage of both these data sources is relatively
high.

\hypertarget{validation-sensitivity-to-data-exclusion}{%
\subsection{Validation: sensitivity to data
exclusion}\label{validation-sensitivity-to-data-exclusion}}

To assess the sensitivity of the cause of death distribution estimates,
we performed a series of validation exercises leaving out certain types
of data. Firstly, we were interested in sensitivity to exclusion of all
data from studies, as these data represented a large amount of
additional information compared to previous efforts (Say et al. 2014).
Additionally, we assessed the change in the estimates when data from
high-income countries were left out. These countries have relatively
large amounts of data available and therefore run the risk of being
overly influential in the hierarchical model set-up. Finally, we
assessed the sensitivity of estimates to an exclusion of 20\% of the
data of each type (CRVS, studies and grey literature).

Table \ref{tab:validation_table} shows the mean absolute difference
between estimates by cause across SDG regions when the model was rerun
excluding any information from studies. Similar results for the other
two exercises (leaving high-income country data out and 20\% of the data
out) are shown in Appendix \ref{appendix_validation}. In general,
differences in estimates are small across all causes and regions,
particularly in Europe and North America, Australia and New Zealand, and
Latin America and the Carribean. In these regions the availability of
CRVS data is high and little information comes from studies. The mean
absolute differences are by far largest in Sub-Saharan Africa, with the
mean difference in embolism, hemorrhage and sepsis proportions being at
least 0.02. The relatively larger differences in Sub-Saharan Africa are
expected, given the lack of other data available in this region means
the data from studies is quite influential.

\begin{table}

\caption{\label{tab:validation_table}Mean absolute difference in estimated country proportions, leaving out observations from studies}
\centering
\resizebox{\linewidth}{!}{
\begin{tabular}[t]{l>{\raggedleft\arraybackslash}p{3cm}>{\raggedleft\arraybackslash}p{3cm}>{\raggedleft\arraybackslash}p{3cm}>{\raggedleft\arraybackslash}p{3cm}>{\raggedleft\arraybackslash}p{3cm}>{\raggedleft\arraybackslash}p{3cm}>{\raggedleft\arraybackslash}p{3cm}>{\raggedleft\arraybackslash}p{3cm}>{}p{3cm}}
\toprule
Cause & Central and Southern Asia & Europe and Northern America & Northern Africa and Western Asia & Oceania excl. Australia and New Zealand & Sub-Saharan Africa & Latin America and the Caribbean & Australia and New Zealand & Eastern and South-Eastern Asia\\
\midrule
\cellcolor{gray!6}{ABO} & \cellcolor{gray!6}{0.009} & \cellcolor{gray!6}{0.001} & \cellcolor{gray!6}{0.001} & \cellcolor{gray!6}{0.007} & \cellcolor{gray!6}{0.007} & \cellcolor{gray!6}{0.001} & \cellcolor{gray!6}{0.001} & \cellcolor{gray!6}{0.004}\\
DIR & 0.007 & 0.002 & 0.003 & 0.008 & 0.016 & 0.003 & 0.002 & 0.009\\
\cellcolor{gray!6}{EMB} & \cellcolor{gray!6}{0.006} & \cellcolor{gray!6}{0.003} & \cellcolor{gray!6}{0.005} & \cellcolor{gray!6}{0.012} & \cellcolor{gray!6}{0.027} & \cellcolor{gray!6}{0.002} & \cellcolor{gray!6}{0.001} & \cellcolor{gray!6}{0.008}\\
HEM & 0.009 & 0.002 & 0.006 & 0.010 & 0.021 & 0.002 & 0.003 & 0.011\\
\cellcolor{gray!6}{SEP} & \cellcolor{gray!6}{0.014} & \cellcolor{gray!6}{0.002} & \cellcolor{gray!6}{0.003} & \cellcolor{gray!6}{0.011} & \cellcolor{gray!6}{0.020} & \cellcolor{gray!6}{0.002} & \cellcolor{gray!6}{0.003} & \cellcolor{gray!6}{0.008}\\
\addlinespace
IND & 0.005 & 0.003 & 0.004 & 0.010 & 0.010 & 0.003 & 0.003 & 0.009\\
\cellcolor{gray!6}{HYP} & \cellcolor{gray!6}{0.008} & \cellcolor{gray!6}{0.002} & \cellcolor{gray!6}{0.003} & \cellcolor{gray!6}{0.009} & \cellcolor{gray!6}{0.017} & \cellcolor{gray!6}{0.002} & \cellcolor{gray!6}{0.001} & \cellcolor{gray!6}{0.007}\\
\bottomrule
\end{tabular}}
\end{table}

\hypertarget{discussion}{%
\section{\texorpdfstring{Discussion
\label{section_discussion}}{Discussion }}\label{discussion}}

In this paper we presented a Bayesian hierarchical framework to estimate
maternal cause of death distributions at the global, regional and
national levels. A total of 14 cause groups are estimated, encompassed
within the 7 main categories of causes of death: abortion, embolism,
hemorrhage, hypertension, sepsis, other direct causes and indirect
causes. The framework allows for data from various sources to be
combined, and accounts for the fact that we usually do not have data on
all cause of death categories. The model pools information across
regions, accounts for correlation between cause groups, and includes
adjustments for data quality and coverage of the observed deaths,
utilizing information on total maternal mortality and all-cause
mortality more broadly. We illustrated the model on three cases of
varying data availability and quality. While our motivation was to
estimate causes of maternal death, the framework could easily be adapted
and applied to estimate cause of death distributions for other key
population sub-groups, such as under-five mortality or premature
mortality.

The model framework has several advantages over previous efforts to
estimate causes of maternal death. Firstly, while previous efforts chose
the one `best data source' available for each country, we account for
data quality in the model, which in turn allows data from multiple
sources for a particular country to be included in the model (Say et al.
2014). Secondly, we account for under-coverage of available data by
weighting the final estimate by the observed number of total maternal
deaths. This means the final estimate for a particular country is
informed not only by the available data but also mortality trends in the
broader region. Finally, the model accounts for missing observations of
causes of death and allows for studies where only a single cause of
death is observed to be included in the estimation process.

There are, however, some limitations to our approach. One of the main
limitations is that we do not adjust for non-representativeness or bias
of data sources. For example, many of the smaller studies that were
included where CRVS data do not exist come from single institutions or a
group of healthcare institutions from residential areas that are
relatively urban. However, women included in these studies may not be
representative of the broader population. In addition, causes of
maternal death that are observed in healthcare facilities are likely to
be different to causes of maternal death that happen in community
centers or if the death occurs at home. While we include a data quality
term in the model, this only adds uncertainty to the data, rather than
adjusting the proportions of a particular cause up or down. In addition,
the re-weighting of estimates based on under-coverage implicitly assumes
the data we have is a representative sample of the broader population,
and also assumes cause of death distributions are similar within
geographical regions. In practice, is it usually very difficult or
impossible to know the types of women captured within a study (in terms
of their age, socioeconomic status, pre-existing conditions, etc) and
how this differs from the broader population, and thus how it could be
post-stratified to be more representative. Future work will aim to
explore options to better account for non-representative samples, and
include adjustments for countries where possible.

A second limitation is that we do not account for the uncertainty in
cause of death classifications. Much of the data that came from studies
had to be translated into one of the cause of death categories from
`free text' descriptions of cause. In some cases it may have not
necessarily been clear which cause of death category observations belong
to (there may be one or two possibilities). In these cases we split
deaths equally based on possible categorizations and did not account for
the relative likelihood or observing one cause over another.

Reducing maternal mortality continues to be an integral part of the
Sustainable Development Goal agenda. The latest report from the MMEIG on
maternal mortality suggests that many countries are not on track to
reach the goal of less than 70 deaths per 100,000 live births by 2030.
In particular, the 10 countries with the highest MMRs in 2017 have seen
a stagnation or slowing of the annual rate of reduction of maternal
mortality, and therefore remain at the greatest risk (WHO et al. 2019).
Understanding the underlying causes of maternal death is necessary to
evaluate progress and aid resource allocation. The importance of
reducing maternal mortality as a public health priority justifies
continued efforts to improve not only estimates but also the data
collected on causes of death.

\hypertarget{funding-acknowledgement}{%
\section*{Funding acknowledgement}\label{funding-acknowledgement}}
\addcontentsline{toc}{section}{Funding acknowledgement}

This work received funding from the UNDP-UNFPA-UNICEF-WHO-World Bank
Special Programme of Research, Development and Research Training in
Human Reproduction (HRP), a cosponsored programme executed by the World
Health Organization (WHO).

\newpage
\appendix

\hypertarget{assignment-of-icd-10-codes-to-cause-of-death-categories}{%
\section{\texorpdfstring{Assignment of ICD-10 codes to cause of death
categories
\label{appendix_icd10}}{Assignment of ICD-10 codes to cause of death categories }}\label{assignment-of-icd-10-codes-to-cause-of-death-categories}}

Table \ref{tab:table_app_1} gives the classification of ICD-10 codes
into the seven main cause of death categories. Where appropriate, the
classification is further broken down by sub-categories.
\begingroup\fontsize{8}{10}\selectfont

\begin{longtable}[t]{>{}cc>{\centering\arraybackslash}p{30em}}
\caption{\label{tab:table_app_1}\label{fig:fig_app2}ICD-10 codes corresponding to the seven main cause categories.}\\
\toprule
Main Cause & Sub-cause & ICD-10 Codes\\
\midrule
\textbf{Abortion} &  & O00, O00.0, O00.1, O00.2, O00.8, O00.9, O01, O01.0, O01.1, O01.9, O02, O02.0, O02.1, O02.8, O02.9, O03, O03.0, O03.1, O03.2, O03.3, O03.4, O03.5, O03.6, O03.7, O03.8, O03.9, O04, O04.0, O04.1, O04.3, O04.5, O04.6, O04.7, O04.8, O04.9, O05, O05.0, O05.1, O05.2, O05.3, O05.4, O05.5, O05.6, O05.7, O05.8, O05.9, O06, O06.0, O06.1, O06.2, O06.3, O06.4, O06.5, O06.6, O06.7, O06.8, O06.9, O07, O07.0, O07.1, O07.2, O07.3, O07.4, O07.5, O07.6, O07.7, O07.8, O07.9\\
\cmidrule{1-3}
\textbf{Embolism} &  & O22, O22.3, O22.5, O22.8, O22.9, O87, O87.1, O87.3, O87.9, O88, O88.0, O88.1, O88.2, O88.3, O88.8, O87.0, O87.2, O87.8\\
\cmidrule{1-3}
 & Ante-partum & O20, O20.0, O20.8, O20.9, O44, O44.1, O45, O45.0, O45.8, O45.9, O46, O46.0, O46.8, O46.9, O71.0\\
\cmidrule{2-3}
 & Intra-partum & O43, O43.2, O67, O67.0, O67.8, O67.9, O71.1, Rupture NOS\\
\cmidrule{2-3}
 & Post-partum & O72, O72.0, O72.1, O72.2, O72.3\\
\cmidrule{2-3}
\multirow{-4}{*}{\centering\arraybackslash \textbf{Hemorrhage}} & Timing Unknown & O71.3, O71.4, O71.7, Hemorrhage NOS\\
\cmidrule{1-3}
\textbf{Hypertension} &  & O11, O12, O12.0, O12.1, O12.2, O13, O14, O14.0, O14.1, O14.2, O14.9, O15, O15.0, O15.1, O15.2, O15.9, O16\\
\cmidrule{1-3}
 & Anaesthesia & O29, O29.0, O29.1, O29.2, O29.3, O29.5, O29.6, O29.8, O29.9, O74, O74.0, O74.1, O74.2, O74.3, O74.4, O74.6, O74.7, O74.8, O74.9, O89, O89.0, O89.1, O89.2, O89.5, O89.6, O89.8, O89.9, O74.5, O89.4\\
\cmidrule{2-3}
 & Obstetric Trauma & O71.2, O71.5, O71.6, O71.8, O71.9\\
\cmidrule{2-3}
 & Obstructed Labor & O33, O33.0, O33.3, O33.4, O33.5, O33.9, O62, O62.0, O62.1, O62.2, O62.3, O62.4, O62.8, O62.9, O63, O63.0, O63.1, O63.2, O63.9, O64, O64.0, O64.1, O64.2, O64.4, O64.5, O64.8, O64.9, O65, O65.1, O65.4, O65.5, O65.9, O66, O66.0, O66.1, O66.2, O66.3, O66.4, O66.9\\
\cmidrule{2-3}
\multirow{-4}{*}{\centering\arraybackslash \textbf{Other Direct Causes}} & Other & O21.1, O24.4, O26.6, O44.0, O73, O73.0, O73.1, O75.4, O75.8, O75.9, O90, O90.0, O90.1, O90.2, O90.3, O90.4, O90.5, O90.8, O90.9, C58, O21, O21.0, O21.9, O22.0, O22.1, O22.2, O25, O26, O26.0, O26.1, O26.3, O26.5, O26.8, O26.9, O28, O28.5, O28.8, O30, O30.0, O30.1, O30.9, O31, O31.2, O31.8, O32, O32.1, O32.2, O32.4, O32.8, O32.9, O34, O34.0, O34.1, O34.2, O34.3, O34.4, O34.5, O34.6, O34.8, O34.9, O35, O35.0, O35.1, O35.5, O35.8, O35.9, O36, O36.0, O36.1, O36.2, O36.3, O36.4, O36.5, O36.6, O36.7, O36.8, O36.9, O40, O41, O41.0, O41.8, O41.9, O42, O42.0, O42.1, O42.2, O42.9, O43.0, O43.1, O43.8, O43.9, O47.0, O47.9, O48, O60, O60.0, O60.1, O60.2, O60.3, O61, O61.0, O61.9, O68, O68.0, O68.1, O68.2, O68.8, O68.9, O69, O69.0, O69.1, O69.2, O69.4, O69.5, O69.8, O69.9, O70, O70.0, O70.1, O70.2, O70.3, O70.9, O75, O75.0, O75.1, O75.2, O75.6, O75.7, O92, O92.2, O92.3, O92.5, O92.6\\
\cmidrule{1-3}
\textbf{Other Indirect Causes} &  & O10, O10.0, O10.1, O10.2, O10.3, O10.4, O10.9, O24, O24.0, O24.1, O24.2, O24.3, O24.9, O98, O98.0, O98.1, O98.3, O98.4, O98.5, O98.6, O98.8, O98.9, O99.0, O99.1, O99.2, O99.3, O99.4, O99.5, O99.6, O99.7, O99.8\\
\cmidrule{1-3}
 & Ante-partum & O23, O23.0, O23.1, O23.2, O23.3, O23.4, O23.5, O23.9, O41.1\\
\cmidrule{2-3}
 & Intra-partum & O75.3\\
\cmidrule{2-3}
 & Post-partum & O85, O86, O86.0, O86.1, O86.2, O86.3, O86.4, O86.8, O91, O91.1, O91.2\\
\cmidrule{2-3}
\multirow{-4}{*}{\centering\arraybackslash \textbf{Sepsis}} & Timing Unknown & A34\\
\bottomrule
\end{longtable}
\endgroup{}

\newpage

\hypertarget{data-adjustments}{%
\section{\texorpdfstring{Data adjustments
\label{appendix_data_adjustments}}{Data adjustments }}\label{data-adjustments}}

\hypertarget{treatment-of-missing-versus-zero-reported-deaths}{%
\subsection{\texorpdfstring{Treatment of missing versus zero-reported
deaths
\label{subsetion_data_zeros}}{Treatment of missing versus zero-reported deaths }}\label{treatment-of-missing-versus-zero-reported-deaths}}

For a number of country-years we encounter missing observations for some
of the main causes. Sometimes however, zero counts are explicitly
reported in the data for a given cause. In the current analysis we
attempt to make a distinction between a death count for which we have no
information, and hence treat as missing, and one where we are implying
that the true count is, indeed, equal to zero. More specifically, for a
particular country-year we consider an observation of zero maternal
deaths for a particular main cause to be a true zero, if the observation
is of data quality type 1 or 2 (quality type distinction is explained in
more detail in Section \ref{subsection_model_dataquality}), if the data
source for that observation has previously observed non-zero maternal
deaths for that main cause, and if the estimated maternal death count
for that country (2019 UN MMEIG estimates) is less than 7. Put
differently, we are willing to assume that in a country with low
estimated maternal mortality and high quality data a zero maternal death
count for a particular cause that has been previously reported is a true
zero.

\hypertarget{hivaids-deaths}{%
\subsection{HIV/AIDS deaths}\label{hivaids-deaths}}

As indicated earlier maternal death due to HIV/AIDS were not part of the
estimation process and HIV/AIDS-inclusive cause of death distributions
were obtained only after incorporating HIV/AIDS estimates produced
elsewhere (see Section \ref{section_hiv} for more details). To exclude
maternal deaths due to HIV/AIDS, observations classified as ICD-10 code
O98.7 were explicitly excluded from the analysis. However, we believe
this exclusion does not fully account for the possible number of
HIV/AIDS deaths present among deaths from indirect (IND) causes in our
data. For instance, HIV/AIDS deaths could be reported among maternal
deaths classified more broadly as ICD-10 Group 7 or could be reported
under ICD-10 code O98. To account for this, we consider each
country-year observation reporting any maternal deaths as ICD-10 Group 7
or using ICD-10 code O98 and adjust that observation's total deaths from
IND causes by subtracting an appropriately scaled MMEIG estimate of
HIV/AIDS maternal deaths for that country-year, where the scaling is
based on a measure of completeness, or coverage, of the maternal deaths
data.

\hypertarget{multi-year-studies}{%
\subsection{Multi-year studies}\label{multi-year-studies}}

A proportion of studies report maternal deaths as aggregate values over
the entire study period. In such cases, where corresponding yearly
maternal deaths are unavailable and the data cannot otherwise be
disaggregated, to prevent such observation from having disproportionate
weight in the model, the total number of maternal deaths reported are
scaled down to a single-year equivalent. This, however, is done only for
those studies where the total number of deaths reported is more than
five times the number of years of the study.

\newpage

\hypertarget{modeling-regions}{%
\section{\texorpdfstring{Modeling regions
\label{appendix_regions}}{Modeling regions }}\label{modeling-regions}}

The organization of countries into regions \(r = 1, \dots, R\) used in
the estimation model follow the classification used by Say et al.
(2014). Table \ref{tab:region_table} shows the countries included in
each of these regions.

\begingroup\fontsize{8}{10}\selectfont

\begin{longtable}[t]{>{\raggedright\arraybackslash}p{1.5in}>{\raggedright\arraybackslash}p{3.5in}}
\caption{\label{tab:region_table}Organization of countries into regions used in the estimation model}\\
\toprule
Model region & Countries\\
\midrule
\cellcolor{gray!6}{Central Asia} & \cellcolor{gray!6}{Kazakhstan; Kyrgyzstan; Tajikistan; Turkmenistan; Uzbekistan}\\
Eastern Asia & China; Democratic People's Republic of Korea; Mongolia; Republic of Korea\\
\cellcolor{gray!6}{Eastern Africa} & \cellcolor{gray!6}{Burundi; Comoros; Djibouti; Eritrea; Ethiopia; Kenya; Madagascar; Malawi; Mauritius; Mozambique; Rwanda; Seychelles; Somalia; South Sudan; Sudan; United Republic of Tanzania; Uganda; Zambia; Zimbabwe}\\
South-Eastern Asia / Oceania & Brunei Darussalam; Cambodia; Fiji; Indonesia; Kiribati; Lao People's Democratic Republic; Malaysia; Micronesia (Federated States of); Myanmar; Papua New Guinea; Philippines; Samoa; Singapore; Solomon Islands; Thailand; Timor-Leste; Tonga; Vanuatu; Viet Nam\\
\cellcolor{gray!6}{Southern Africa} & \cellcolor{gray!6}{Botswana; Lesotho; Namibia; South Africa; Eswatini}\\
\addlinespace
Western Africa & Benin; Burkina Faso; Cabo Verde; Côte d'Ivoire; Gambia; Ghana; Guinea; Guinea-Bissau; Liberia; Mali; Mauritania; Niger; Nigeria; Senegal; Sierra Leone; Togo\\
\cellcolor{gray!6}{Central America} & \cellcolor{gray!6}{Belize; Costa Rica; El Salvador; Guatemala; Honduras; Mexico; Nicaragua; Panama}\\
Developed regions & Australia; Austria; Belarus; Belgium; Bulgaria; Canada; Czechia; Denmark; Estonia; Finland; France; Germany; Greece; Hungary; Iceland; Ireland; Italy; Japan; Latvia; Lithuania; Luxembourg; Malta; Netherlands; New Zealand; Norway; Poland; Portugal; Republic of Moldova; Romania; Russian Federation; Slovakia; Slovenia; Spain; Sweden; Switzerland; Ukraine; United Kingdom of Great Britain and Northern Ireland; United States of America\\
\cellcolor{gray!6}{Western Asia} & \cellcolor{gray!6}{Armenia; Azerbaijan; Bahrain; Cyprus; Georgia; Iraq; Israel; Jordan; Kuwait; Lebanon; West Bank and Gaza Strip; Oman; Qatar; Saudi Arabia; Syrian Arab Republic; Turkey; United Arab Emirates; Yemen}\\
South America & Argentina; Bolivia (Plurinational State of); Brazil; Chile; Colombia; Ecuador; Guyana; Paraguay; Peru; Suriname; Uruguay; Venezuela (Bolivarian Republic of)\\
\addlinespace
\cellcolor{gray!6}{Caribbean} & \cellcolor{gray!6}{Antigua and Barbuda; Bahamas; Barbados; Cuba; Dominican Republic; Grenada; Haiti; Jamaica; Puerto Rico; Saint Lucia; Saint Vincent and the Grenadines; Trinidad and Tobago}\\
Middle Africa & Angola; Cameroon; Central African Republic; Chad; Congo; Democratic Republic of the Congo; Equatorial Guinea; Gabon; Sao Tome and Principe\\
\cellcolor{gray!6}{Northern Africa} & \cellcolor{gray!6}{Algeria; Egypt; State of Libya; Morocco; Tunisia}\\
Transition countries of South-Eastern Europe & Albania; Bosnia and Herzegovina; Croatia; Montenegro; Serbia; Republic of North Macedonia\\
\cellcolor{gray!6}{Southern Asia} & \cellcolor{gray!6}{Afghanistan; Bangladesh; Bhutan; India; Iran (Islamic Republic of); Maldives; Nepal; Pakistan; Sri Lanka}\\
\bottomrule
\end{longtable}
\endgroup{}

\newpage

\hypertarget{additional-validation-results}{%
\section{\texorpdfstring{Additional validation results
\label{appendix_validation}}{Additional validation results }}\label{additional-validation-results}}

Tables \ref{tab:dev_table} and \ref{tab:l20o_table} show the mean
absolute differences between estimates by region and main cause of death
when excluding all data from high-income countries, and 20\% of the data
from each data quality type (1, 2, 3, 4) respectively. Removing data
from high-income countries has a relatively large affect on estimates in
Europe and North America, and Australia and New Zealand, but minimal
effects in all other regions. This suggests that high-income regions,
that have large amounts of data available, are not overly influential on
estimates in other regions. Leaving out 20\% of the available data by
type at random has minimal effect on estimates across all regions and
cause groups.

\begin{table}

\caption{\label{tab:dev_table}Mean absolute difference in estimated country proportions, leaving out observations from 'Developed regions'}
\centering
\resizebox{\linewidth}{!}{
\begin{tabular}[t]{l>{\raggedleft\arraybackslash}p{3cm}>{\raggedleft\arraybackslash}p{3cm}>{\raggedleft\arraybackslash}p{3cm}>{\raggedleft\arraybackslash}p{3cm}>{\raggedleft\arraybackslash}p{3cm}>{\raggedleft\arraybackslash}p{3cm}>{\raggedleft\arraybackslash}p{3cm}>{\raggedleft\arraybackslash}p{3cm}>{}p{3cm}}
\toprule
Cause & Central and Southern Asia & Europe and Northern America & Northern Africa and Western Asia & Oceania excl. Australia and New Zealand & Sub-Saharan Africa & Latin America and the Caribbean & Australia and New Zealand & Eastern and South-Eastern Asia\\
\midrule
\cellcolor{gray!6}{ABO} & \cellcolor{gray!6}{0.001} & \cellcolor{gray!6}{0.012} & \cellcolor{gray!6}{0.001} & \cellcolor{gray!6}{0.001} & \cellcolor{gray!6}{0.001} & \cellcolor{gray!6}{0.001} & \cellcolor{gray!6}{0.033} & \cellcolor{gray!6}{0.002}\\
DIR & 0.001 & 0.047 & 0.002 & 0.001 & 0.001 & 0.002 & 0.055 & 0.004\\
\cellcolor{gray!6}{EMB} & \cellcolor{gray!6}{0.004} & \cellcolor{gray!6}{0.116} & \cellcolor{gray!6}{0.008} & \cellcolor{gray!6}{0.003} & \cellcolor{gray!6}{0.002} & \cellcolor{gray!6}{0.003} & \cellcolor{gray!6}{0.139} & \cellcolor{gray!6}{0.013}\\
HEM & 0.003 & 0.045 & 0.003 & 0.008 & 0.004 & 0.002 & 0.103 & 0.008\\
\cellcolor{gray!6}{SEP} & \cellcolor{gray!6}{0.002} & \cellcolor{gray!6}{0.018} & \cellcolor{gray!6}{0.002} & \cellcolor{gray!6}{0.001} & \cellcolor{gray!6}{0.001} & \cellcolor{gray!6}{0.002} & \cellcolor{gray!6}{0.005} & \cellcolor{gray!6}{0.004}\\
\addlinespace
IND & 0.003 & 0.070 & 0.003 & 0.007 & 0.003 & 0.004 & 0.177 & 0.009\\
\cellcolor{gray!6}{HYP} & \cellcolor{gray!6}{0.004} & \cellcolor{gray!6}{0.040} & \cellcolor{gray!6}{0.006} & \cellcolor{gray!6}{0.005} & \cellcolor{gray!6}{0.007} & \cellcolor{gray!6}{0.002} & \cellcolor{gray!6}{0.089} & \cellcolor{gray!6}{0.008}\\
\bottomrule
\end{tabular}}
\end{table}

\begin{table}

\caption{\label{tab:l20o_table}Mean absolute difference in estimated country proportions, leaving out 20 percent of observations from each data quality type.}
\centering
\resizebox{\linewidth}{!}{
\begin{tabular}[t]{l>{\raggedleft\arraybackslash}p{3cm}>{\raggedleft\arraybackslash}p{3cm}>{\raggedleft\arraybackslash}p{3cm}>{\raggedleft\arraybackslash}p{3cm}>{\raggedleft\arraybackslash}p{3cm}>{\raggedleft\arraybackslash}p{3cm}>{\raggedleft\arraybackslash}p{3cm}>{\raggedleft\arraybackslash}p{3cm}>{}p{3cm}}
\toprule
Cause & Central and Southern Asia & Europe and Northern America & Northern Africa and Western Asia & Oceania excl. Australia and New Zealand & Sub-Saharan Africa & Latin America and the Caribbean & Australia and New Zealand & Eastern and South-Eastern Asia\\
\midrule
\cellcolor{gray!6}{ABO} & \cellcolor{gray!6}{0.002} & \cellcolor{gray!6}{0.003} & \cellcolor{gray!6}{0.001} & \cellcolor{gray!6}{0.009} & \cellcolor{gray!6}{0.002} & \cellcolor{gray!6}{0.008} & \cellcolor{gray!6}{0.001} & \cellcolor{gray!6}{0.006}\\
DIR & 0.005 & 0.010 & 0.004 & 0.004 & 0.002 & 0.007 & 0.004 & 0.004\\
\cellcolor{gray!6}{EMB} & \cellcolor{gray!6}{0.003} & \cellcolor{gray!6}{0.028} & \cellcolor{gray!6}{0.003} & \cellcolor{gray!6}{0.001} & \cellcolor{gray!6}{0.002} & \cellcolor{gray!6}{0.004} & \cellcolor{gray!6}{0.017} & \cellcolor{gray!6}{0.003}\\
HEM & 0.004 & 0.011 & 0.004 & 0.004 & 0.006 & 0.006 & 0.017 & 0.005\\
\cellcolor{gray!6}{SEP} & \cellcolor{gray!6}{0.003} & \cellcolor{gray!6}{0.005} & \cellcolor{gray!6}{0.003} & \cellcolor{gray!6}{0.005} & \cellcolor{gray!6}{0.003} & \cellcolor{gray!6}{0.003} & \cellcolor{gray!6}{0.009} & \cellcolor{gray!6}{0.006}\\
\addlinespace
IND & 0.004 & 0.021 & 0.008 & 0.009 & 0.007 & 0.009 & 0.008 & 0.010\\
\cellcolor{gray!6}{HYP} & \cellcolor{gray!6}{0.002} & \cellcolor{gray!6}{0.008} & \cellcolor{gray!6}{0.003} & \cellcolor{gray!6}{0.005} & \cellcolor{gray!6}{0.001} & \cellcolor{gray!6}{0.006} & \cellcolor{gray!6}{0.010} & \cellcolor{gray!6}{0.003}\\
\bottomrule
\end{tabular}}
\end{table}

\newpage

\hypertarget{references}{%
\section*{References}\label{references}}
\addcontentsline{toc}{section}{References}

\hypertarget{refs}{}
\begin{CSLReferences}{1}{0}
\leavevmode\vadjust pre{\hypertarget{ref-abouchadi2018underreporting}{}}%
Abouchadi, Saloua, Wei-Hong Zhang, and Vincent De Brouwere. 2018.
{``Underreporting of Deaths in the Maternal Deaths Surveillance System
in One Region of Morocco.''} \emph{PloS One} 13 (1): e0188070.

\leavevmode\vadjust pre{\hypertarget{ref-alexander2018global}{}}%
Alexander, Monica, and Leontine Alkema. 2018. {``Global Estimation of
Neonatal Mortality Using a Bayesian Hierarchical Splines Regression
Model.''} \emph{Demographic Research} 38: 335--72.

\leavevmode\vadjust pre{\hypertarget{ref-alkema2016global}{}}%
Alkema, Leontine, Doris Chou, Daniel Hogan, Sanqian Zhang, Ann-Beth
Moller, Alison Gemmill, Doris Ma Fat, et al. 2016. {``Global, Regional,
and National Levels and Trends in Maternal Mortality Between 1990 and
2015, with Scenario-Based Projections to 2030: A Systematic Analysis by
the UN Maternal Mortality Estimation Inter-Agency Group.''} \emph{The
Lancet} 387 (10017): 462--74.

\leavevmode\vadjust pre{\hypertarget{ref-alkema2014global}{}}%
Alkema, Leontine, and Jin Rou New. 2014. {``Global Estimation of Child
Mortality Using a Bayesian b-Spline Bias-Reduction Model.''} \emph{The
Annals of Applied Statistics}, 2122--49.

\leavevmode\vadjust pre{\hypertarget{ref-alkema2017bayesian}{}}%
Alkema, Leontine, Sanqian Zhang, Doris Chou, Alison Gemmill, Ann-Beth
Moller, Doris Ma Fat, Lale Say, Colin Mathers, Daniel Hogan, and others.
2017. {``A Bayesian Approach to the Global Estimation of Maternal
Mortality.''} \emph{The Annals of Applied Statistics} 11 (3): 1245--74.

\leavevmode\vadjust pre{\hypertarget{ref-alvarez2009factors}{}}%
Alvarez, Jose Luis, Ruth Gil, Valentin Hernandez, and Angel Gil. 2009.
{``Factors Associated with Maternal Mortality in Sub-Saharan Africa: An
Ecological Study.''} \emph{BMC Public Health} 9 (1): 462.

\leavevmode\vadjust pre{\hypertarget{ref-briozzo2016overall}{}}%
Briozzo, Leonel, Rodolfo Gómez Ponce de León, Giselle Tomasso, and
Anibal Faúndes. 2016. {``Overall and Abortion-Related Maternal Mortality
Rates in Uruguay over the Past 25 Years and Their Association with
Policies and Actions Aimed at Protecting Women's Rights.''}
\emph{International Journal of Gynecology \& Obstetrics} 134: S20--23.

\leavevmode\vadjust pre{\hypertarget{ref-cahill2018modern}{}}%
Cahill, Niamh, Emily Sonneveldt, John Stover, Michelle Weinberger,
Jessica Williamson, Chuchu Wei, Win Brown, and Leontine Alkema. 2018.
{``Modern Contraceptive Use, Unmet Need, and Demand Satisfied Among
Women of Reproductive Age Who Are Married or in a Union in the Focus
Countries of the Family Planning 2020 Initiative: A Systematic Analysis
Using the Family Planning Estimation Tool.''} \emph{The Lancet} 391
(10123): 870--82.

\leavevmode\vadjust pre{\hypertarget{ref-cdc2015}{}}%
CDC. 2015. {``Global Program for Civil Registration and Vital Statistics
(CRVS) Improvement.''} Available at:
\url{https://www.cdc.gov/nchs/isp/isp_crvs.htm}.

\leavevmode\vadjust pre{\hypertarget{ref-eriksson2013accuracy}{}}%
Eriksson, Anders, Hans Stenlund, Kristin Ahlm, Kurt Boman, Lars Olov
Bygren, Lars Age Johansson, Bert-Ove Olofsson, Stig Wall, and Lars
Weinehall. 2013. {``Accuracy of Death Certificates of Cardiovascular
Disease in a Community Intervention in Sweden.''} \emph{Scandinavian
Journal of Public Health} 41 (8): 883--89.

\leavevmode\vadjust pre{\hypertarget{ref-gabry2020cmdstanr}{}}%
Gabry, Jonah, and Rok Češnovar. 2020. \emph{Cmdstanr: R Interface to
'CmdStan'}.

\leavevmode\vadjust pre{\hypertarget{ref-gerdts2015measuring}{}}%
Gerdts, Caitlin, Ozge Tunçalp, Heidi Johnston, and Bela Ganatra. 2015.
{``Measuring Abortion-Related Mortality: Challenges and
Opportunities.''} \emph{Reproductive Health} 12 (1): 1--3.

\leavevmode\vadjust pre{\hypertarget{ref-gerdts2013measuring}{}}%
Gerdts, Caitlin, Divya Vohra, and Jennifer Ahern. 2013. {``Measuring
Unsafe Abortion-Related Mortality: A Systematic Review of the Existing
Methods.''} \emph{PloS One} 8 (1): e53346.

\leavevmode\vadjust pre{\hypertarget{ref-ghosh2006equivalence}{}}%
Ghosh, Malay, Li Zhang, and Bhramar Mukherjee. 2006. {``Equivalence of
Posteriors in the Bayesian Analysis of the Multinomial-Poisson
Transformation.''} \emph{Metron - International Journal of Statistics}
LXIV (February): 19--28.

\leavevmode\vadjust pre{\hypertarget{ref-lewandowski2009generating}{}}%
Lewandowski, Daniel, Dorota Kurowicka, and Harry Joe. 2009.
{``Generating Random Correlation Matrices Based on Vines and Extended
Onion Method.''} \emph{Journal of Multivariate Analysis} 100 (9):
1989--2001.

\leavevmode\vadjust pre{\hypertarget{ref-liu2016global}{}}%
Liu, Li, Shefali Oza, Dan Hogan, Yue Chu, Jamie Perin, Jun Zhu, Joy E
Lawn, Simon Cousens, Colin Mathers, and Robert E Black. 2016. {``Global,
Regional, and National Causes of Under-5 Mortality in 2000--15: An
Updated Systematic Analysis with Implications for the Sustainable
Development Goals.''} \emph{The Lancet} 388 (10063): 3027--35.

\leavevmode\vadjust pre{\hypertarget{ref-macdorman2016united}{}}%
MacDorman, Marian F, Eugene Declercq, Howard Cabral, and Christine
Morton. 2016. {``Is the United States Maternal Mortality Rate
Increasing? Disentangling Trends from Measurement Issues Short Title: US
Maternal Mortality Trends.''} \emph{Obstetrics and Gynecology} 128 (3):
447.

\leavevmode\vadjust pre{\hypertarget{ref-maduka2020preventing}{}}%
Maduka, Omosivie, and Rosemary Ogu. 2020. {``Preventing Maternal
Mortality During Childbirth: The Scourge of Delivery with Unskilled
Birth Attendants.''} In \emph{Childbirth}. IntechOpen.

\leavevmode\vadjust pre{\hypertarget{ref-mbizvo2012global}{}}%
Mbizvo, Michael T, and Lale Say. 2012. {``Global Progress and
Potentially Effective Policy Responses to Reduce Maternal Mortality.''}
\emph{International Journal of Gynecology \& Obstetrics} 119: S9--12.

\leavevmode\vadjust pre{\hypertarget{ref-mcalister2006female}{}}%
McAlister, Chryssa, and Thomas F Baskett. 2006. {``Female Education and
Maternal Mortality: A Worldwide Survey.''} \emph{Journal of Obstetrics
and Gynaecology Canada} 28 (11): 983--90.

\leavevmode\vadjust pre{\hypertarget{ref-messite1996accuracy}{}}%
Messite, Jacqueline, and Steven D Stellman. 1996. {``Accuracy of Death
Certificate Completion: The Need for Formalized Physician Training.''}
\emph{Jama} 275 (10): 794--96.

\leavevmode\vadjust pre{\hypertarget{ref-montgomery2014effect}{}}%
Montgomery, Ann L, Shaza Fadel, Rajesh Kumar, Sue Bondy, Rahim
Moineddin, and Prabhat Jha. 2014. {``The Effect of Health-Facility
Admission and Skilled Birth Attendant Coverage on Maternal Survival in
India: A Case-Control Analysis.''} \emph{PloS One} 9 (6): e95696.

\leavevmode\vadjust pre{\hypertarget{ref-muldoon2011health}{}}%
Muldoon, Katherine A, Lindsay P Galway, Maya Nakajima, Steve Kanters,
Robert S Hogg, Eran Bendavid, and Edward J Mills. 2011. {``Health System
Determinants of Infant, Child and Maternal Mortality: A Cross-Sectional
Study of UN Member Countries.''} \emph{Globalization and Health} 7 (1):
42.

\leavevmode\vadjust pre{\hypertarget{ref-naghavi2017global}{}}%
Naghavi, Mohsen, Amanuel Alemu Abajobir, Cristiana Abbafati, Kaja M
Abbas, Foad Abd-Allah, Semaw Ferede Abera, Victor Aboyans, et al. 2017.
{``Global, Regional, and National Age-Sex Specific Mortality for 264
Causes of Death, 1980--2016: A Systematic Analysis for the Global Burden
of Disease Study 2016.''} \emph{The Lancet} 390 (10100): 1151--1210.

\leavevmode\vadjust pre{\hypertarget{ref-nair2017indirect}{}}%
Nair, Manisha, Catherine Nelson-Piercy, and Marian Knight. 2017.
{``Indirect Maternal Deaths: UK and Global Perspectives.''}
\emph{Obstetric Medicine} 10 (1): 10--15.

\leavevmode\vadjust pre{\hypertarget{ref-prata2011there}{}}%
Prata, Ndola, Paige Passano, Tami Rowen, Suzanne Bell, Julia Walsh, and
Malcolm Potts. 2011. {``Where There Are (Few) Skilled Birth
Attendants.''} \emph{Journal of Health, Population, and Nutrition} 29
(2): 81.

\leavevmode\vadjust pre{\hypertarget{ref-raftery2014bayesian}{}}%
Raftery, Adrian E, Leontine Alkema, and Patrick Gerland. 2014.
{``Bayesian Population Projections for the United Nations.''}
\emph{Statistical Science: A Review Journal of the Institute of
Mathematical Statistics} 29 (1): 58.

\leavevmode\vadjust pre{\hypertarget{ref-salanave1999classification}{}}%
Salanave, Benoit, Marie-Helene Bouvier-Colle, Noelle Varnoux, Sophie
Alexander, and Alison Macfarlane. 1999. {``Classification Differences
and Maternal Mortality: A European Study. MOMS Group. MOthers' Mortality
and Severe Morbidity.''} \emph{International Journal of Epidemiology} 28
(1): 64--69.

\leavevmode\vadjust pre{\hypertarget{ref-say2014global}{}}%
Say, Lale, Doris Chou, Alison Gemmill, Özge Tunçalp, Ann-Beth Moller,
Jane Daniels, A Metin Gülmezoglu, Marleen Temmerman, and Leontine
Alkema. 2014. {``Global Causes of Maternal Death: A WHO Systematic
Analysis.''} \emph{The Lancet Global Health} 2 (6): e323--33.

\leavevmode\vadjust pre{\hypertarget{ref-schumacher2020flexible}{}}%
Schumacher, Austin E, Tyler H McCormick, Jon Wakefield, Yue Chu, Jamie
Perin, Francisco Villavicencio, Noah Simon, and Li Liu. 2020. {``A
Flexible Bayesian Framework to Estimate Age-and Cause-Specific Child
Mortality over Time from Sample Registration Data.''} \emph{arXiv
Preprint arXiv:2003.00401}.

\leavevmode\vadjust pre{\hypertarget{ref-stan2019stan}{}}%
Stan Development Team. 2019. {``Stan Modeling Language Users Guide and
Reference Manual Version 2.25.''} \url{https://mc-stan.org/}.

\leavevmode\vadjust pre{\hypertarget{ref-unaids2017}{}}%
UNAIDS. 2017. {``UNAIDS Data 2017.''} Available at:
\url{https://www.unaids.org/en/resources/documents/2017/2017_data_book}.

\leavevmode\vadjust pre{\hypertarget{ref-sdg2020}{}}%
UNDESA. 2020. {``SDG 3: Ensure Healthy Lives and Promote Well-Being for
All at All Ages.''} Available at: \url{https://sdgs.un.org/goals/goal3}.

\leavevmode\vadjust pre{\hypertarget{ref-unpd2019}{}}%
UNPD. 2019. {``World Population Prospects: The 2019 Edition.''}
Available at: \url{http://esa.un.org/wpp/}.

\leavevmode\vadjust pre{\hypertarget{ref-wakefield2019estimating}{}}%
Wakefield, Jon, Geir-Arne Fuglstad, Andrea Riebler, Jessica Godwin,
Katie Wilson, and Samuel J Clark. 2019. {``Estimating Under-Five
Mortality in Space and Time in a Developing World Context.''}
\emph{Statistical Methods in Medical Research} 28 (9): 2614--34.

\leavevmode\vadjust pre{\hypertarget{ref-wang2017global}{}}%
Wang, Haidong, Amanuel Alemu Abajobir, Kalkidan Hassen Abate, Cristiana
Abbafati, Kaja M Abbas, Foad Abd-Allah, Semaw Ferede Abera, et al. 2017.
{``Global, Regional, and National Under-5 Mortality, Adult Mortality,
Age-Specific Mortality, and Life Expectancy, 1970--2016: A Systematic
Analysis for the Global Burden of Disease Study 2016.''} \emph{The
Lancet} 390 (10100): 1084--1150.

\leavevmode\vadjust pre{\hypertarget{ref-who2012}{}}%
WHO. 2012. {``The {WHO} Application of {ICD-10} to Deaths During
Pregnancy, Childbirth and Puerperium: {ICD MM}.''} Available at:
\url{https://www.who.int/reproductivehealth/publications/monitoring/9789241548458/en/}.

\leavevmode\vadjust pre{\hypertarget{ref-gho2020}{}}%
---------. 2020. {``The Global Health Observatory: Maternal Deaths.''}
Available at:
\url{https://www.who.int/data/gho/indicator-metadata-registry/imr-details/4622}.

\leavevmode\vadjust pre{\hypertarget{ref-who2019}{}}%
WHO, UNICEF, UNFPA, World Bank Group, and United Nations Population
Division. 2019. {``Maternal Mortality: Levels and Trends 2000 to
2017.''} Available at:
\url{https://www.who.int/reproductivehealth/publications/maternal-mortality-2000-2017/en/}.

\end{CSLReferences}

\end{document}